\documentclass[prl,superscriptaddress,twocolumn]{revtex4-2}
\usepackage{graphics,graphicx, array, afterpage}
\usepackage{qcircuit,amsmath}
\usepackage{grffile}
\usepackage[export]{adjustbox}
\usepackage{lineno}
\usepackage{xcolor}
\usepackage{longtable}
\usepackage{braket}
\usepackage{array}
\usepackage{multirow}
\usepackage{enumerate}
\usepackage{amsmath}
\usepackage{amssymb}
\usepackage{esint}
\usepackage{amsthm}
\usepackage{times}
\usepackage{multirow}
\usepackage[colorlinks, citecolor=blue]{hyperref}
\hypersetup{linkcolor=magenta,urlcolor=blue,citecolor=blue,pdfstartview={FitH},urlcolor=blue}

\newcommand{\tr}[1]{\textnormal{Tr}[#1]}

\begin{document}
\title{Non-Bloch self-energy of dissipative interacting fermions}

\author{He-Ran Wang}
\thanks{These authors contributed equally to this work.}
\affiliation{Institute for Advanced Study, Tsinghua University, Beijing 100084,
People's Republic of China}

\author{Zijian Wang}
\thanks{These authors contributed equally to this work.}
\affiliation{Institute for Advanced Study, Tsinghua University, Beijing 100084,
People's Republic of China}

\author{Zhong Wang}
\email{wangzhongemail@tsinghua.edu.cn}
\affiliation{Institute for Advanced Study, Tsinghua University, Beijing 100084,
People's Republic of China}

\begin{abstract}

The non-Hermitian skin effect describes the phenomenon of exponential localization of single-particle eigenstates near the boundary of the system. 
We consider its generalization to the many-body regime by investigating a general class of interacting fermion lattice models in Markovian open quantum systems.
Therein, the elementary excitations from the ``vacuum'' (steady state) are given by two types of dissipative fermionic modes composed of single-fermion operators, which govern the long-time nonequilibrium dynamics.
We perturbatively calculate the self-energy matrix of these bare modes in the presence of interactions, and utilize the non-Bloch band theory to derive an exact integral representation.
By imposing complex momentum conservation, we obtain a simplified expression for corrections to Liouvillian spectrum that agrees well with numerical calculations to high precision. 
We further perform perturbative analysis of Liouvillian eigenstates and identify signatures of interaction-enhanced NHSE at the quasiparticle level, manifested as renormalization of the generalized Brillouin zone.
Our results establish a diagrammatic framework for dissipative interacting fermions with non-Hermitian skin effect in a description of full-fledged Lindblad master equations, which resembles Fermi liquid theory in terms of interaction-dressed quasiparticles.
\end{abstract}

\maketitle

\textit{Introduction}---When a quantum system interacts with the environment, an accurate description of its dynamics requires incorporating non-Hermiticity into the unitary evolution. In recent years, intriguing phenomena have been discovered in open quantum systems and non-Hermitian models, among which the non-Hermitian skin effect (NHSE) has attracted considerable attention~\cite{yao2018edge,kunst2018biorthogonal, Alvarez2018Robust, Lee2019Anatomy, Xiao2020Non, Helbig2020Generalized, Ghatak2020Observation, Wang2022morphing, Ashida2021Non, Wang2024Tutorial, Gohsrich2024Perspective}. The hallmark of NHSE is the unusual behavior of eigenstates, that they accumulate exponentially near the boundary under the open boundary condition (OBC), in contrast to the homogeneous distribution along the chain under periodic boundary condition (PBC). 
Despite the explicit breaking of translational invariance, the eigenstates can be labeled by complex momentum with nonzero imaginary part, according to the recently established non-Bloch band theory~\cite{yao2018edge, Yokomizo2019, Longhi2019Probing,Song2019Topological, Kawabata2020Symplectic, Longhi2020Collapse, yang2020Auxiliary, Lee2020Unraveling, Yi2020Onsite, xue2021simple, Hu2023Geometric, Li2024Classical}.
All the complex momenta of eigenstates form closed curves on the complex plane, which is dubbed the generalized Brillouin zone (GBZ)~\cite{yao2018edge, Yokomizo2019, yang2020Auxiliary, Zhang2020Correspondence}. 

The extensions of single-particle NHSE to the many-body regime could give rise to novel phases of matter~\cite{ yang2021exceptional, suthar2022non, Hu2023Many, Longhi2023Spectral, Hamanaka2024Interaction, Gliozzi2024Manybody, Qin2024Occupation, Gliozzi2024Many, Yoshida2024Mott,Brighi2024Nonreciprocal,mu2020emergent, dora2022full}. 
The characterization of many-body topological phases in the presence of NHSE requires proper generalizations of closed-system topological indicators~\cite{lee2020many, liu2020non, yoshida2022reduction,alsallom2022fate, kawabata2022many, chen2022characterizing, Zhang2022Symmetry, Kim2024Collective}. 
On the other hand, many-body interactions can become the source of NHSE (e.g. through non-Hermitian self-energy~\cite{yi2020non, okuma2021non, yoshida2021real, geng2022nonreciprocal, Shin2023Z2, Micallo2023Correlation}, integrable inter-particle scattering~\cite{Wang2023Scale, Guo2023Boost, Zheng2024Exact}, or density-assisted non-reciprocal hoppings~\cite{Lee2021Many, Shen2022Non, Faugno2022Interaction, Li2023Many}). 
Various analytical and numerical techniques have been introduced to address the many-body problem in this context, including non-Hermitian Bethe ansatz~\cite{Wang2023Scale, Guo2023Boost, Mao2023Non, Zheng2024Exact, Wang2026Explicit} and tensor network techniques~\cite{alsallom2022fate, Chen2023Topo, Begg2024Quantum, Brighi2024Nonreciprocal}. 
However, prior work has primarily focused on effective non-Hermitian Hamiltonian descriptions of open quantum systems by neglecting quantum jumps, of which practical realizations require postselections and generally can lead to exponential overhead.
For full-fledged open quantum systems, a systematic understanding of the interplay between interactions and NHSE remains elusive, particularly regarding the role of the GBZ in many-body physics.

In this paper, we make progress on this subject by developing a GBZ-based diagrammatic perturbation theory for interacting fermionic chains in contact with Markovian reservoirs. The dynamics of the system density matrix $\rho$ is governed by the Lindblad master equation~\cite{breuer2002theory}:
\begin{equation}
	\frac{\mathrm{d}\rho}{\mathrm{d}t}=-i[H,\rho]+\sum_{\mu}(2L_{\mu}\rho L_{\mu}^{\dagger}-\{L_{\mu}^{\dagger}L_{\mu},\rho\})\equiv\mathcal{L}\rho,  
\end{equation}
where $H$ is the system Hamiltonian, $L_\mu$ are the jump operators with index $\mu$ running over all channels coupled to reserviors, and $\mathcal{L}$ denotes the Liouvillian superoperator. 
For the non-interacting counterpart, where $\mathcal{L}$ is quadratic in fermionic operators, it has been shown that NHSE can emerge through dissipative engineering, manifesting as anomalous boundary-dependent damping of observables~\cite{Song2019Chiral, Haga2021Slowing, Yang2022Exactly, Lee2023Anomalously}. In that setting, the analytical tractability relies on the closure of the equations of motion for single-particle correlation functions, which are governed by a ``damping matrix'' that formally resembles a single-particle non-Hermitian tight-binding model \cite{Song2019Chiral,mcdonald2022non}.

Here, we incorporate density-density interactions, which render the correlation-function approach impractical. Instead, we adopt the vectorization method to map the full Liouvillian superoperator to a non-Hermitian many-body Hamiltonian acting on the doubled Hilbert space~\cite{schmutz1978real, Prosen2008Third, Seligman2010Third, Barthel2022Solving,McDonald2023Third}. Within this framework, we identify a quasiparticle description in which modes are labeled by complex momenta on GBZ and interact via density-assisted pairing interactions. In this picture, the vacuum corresponds to the physical steady state, while excitations correspond to the dissipative eigenmodes of the Liouvillian. To treat the pairing interactions, we develop a diagrammatic perturbation theory in which interactions enter through self-energy corrections to the damping matrix.

In contrast to conventional diagrammatic techniques developed for open-system field theory and lattice models~\cite{Altland2010Condensed, Kamenev2011Field, Sieberer2016Keldysh}, a crucial distinction brought by NHSE is that propagators in Feynman diagrams carry complex momenta on GBZ without usual momentum conservation laws \textit{a priori}, motivating the term ``non-Bloch self-energy''. This also poses a technical challenge: the self-energy formula involves triple integrals over GBZ, which generally lacks closed-form expressions. We overcome this obstruction by deforming integral contours and effectively restoring momentum conservation, which yields a simplified self-energy formula for Liouvillian eigenvalues as double integrals over the Brillouin zone (BZ).

Using a Liouvillian version of the Hatano-Nelson model as a benchmark, we validate our approach through comparison with numerical results. We further perform perturbative calculations of eigenstates and GBZ, and show the enhancement of NHSE. The validity of perturbation theory in one-dimensional fermionic open systems is also discussed and elucidated.
Our formulation can be viewed as an open-system analogue of Fermi-liquid theory, in the sense that interactions dress the bare modes into quasiparticles within a controlled diagrammatic framework.

\textit{Model and setup}---We consider fermionic Liouvillian $\mathcal{L}=\mathcal{L}_0 + \mathcal{L}_I$, where $ \mathcal{L}_0$ is quadratic in fermionic operators, and $\mathcal{L}_I$ accounts for interactions.
$\mathcal{L}_0$ consists of a quadratic Hamiltonian $H=\sum_{ij}h_{ij} c_i^\dagger c_j$ and quantum jump operators being linear in fermionic operators: $L_{\mu}^l=\sum_i D_{\mu i}^l c_i,~L_{\mu}^g=\sum_i D_{\mu i }^g c_i^\dagger$, where $c_i^\dagger(c_i),~i=1,\cdots, N$ are fermionic creation (annihilation) operators on a 1D lattice. 
$\mathcal{L}_I$ is taken to be the density-density interacting Hamiltonian of finite range: $H_I=\sum_{ i\le j}U_{ij}n_in_j$, $n_i = c_i^\dagger c_i$. The correlated quantum jump operators can be implemented through reservoir engineering on various experimental platforms, including ultracold atoms~\cite{Gou2020Tunable, Liang2022Dynamic, Zhao2025Two} and quantum dots~\cite{Malz2018Current}, both of which naturally admit interactions.

First, we introduce the vectorization method for fermionic open systems.
The idea is to map the density matrix to a state vector in the doubled Hilbert space: $\rho=\sum_{m,n}\rho_{mn}\ket{m}\bra{n}$$\to$$\ket{\rho}=\sum_{m,n}\rho_{mn}\ket{m}\otimes\ket{n}$,which carries both the left and right indices of the density matrix. We choose basis states as fermion number eigenstates:$|m\rangle=|m_1,\cdots,m_N\rangle,|n\rangle=|\bar{n}_1,\cdots,\bar{n}_N\rangle$, where $m_i(\bar{n}_i)$ denotes the occupation on site $i$.
Accordingly, operators acting on $\rho$ from the left and right of $\rho$ should be distinguished, and we denote right actions of an operator $O$ by a bar: $\bar{O}$.
Consequently, the quadratic Liouvillian $\mathcal{L}_0$ admits the form \cite{schmutz1978real,Prosen2008Third,Seligman2010Third,Barthel2022Solving,McDonald2023Third}:
 \begin{align}\label{FreeLind}
     \mathcal{L}_0=&\boldsymbol{c}^\dagger\mathbb{L}_0\boldsymbol{c}-\tr{M^l+(M^g)^T-ih},\nonumber\\
\mathbb{L}_0=&\left(\begin{array}{cc}
	-ih+(M^g)^T-M^l & 2(M^g)^T \\
	2M^l& -ih-(M^g)^T+M^l
\end{array} \right).
 \end{align}
Here, we introduce positive-semidefinite matrices arising from quantum jump operators:
$M^{g(l)}_{ij}=\sum_\mu D^{g(l)*}_{\mu i}D^{g(l)}_{\mu j}$, and the Nambu basis of left and right fermions: $\boldsymbol{c}=(\eta c_1,\cdots, \eta c_N,\bar{c}_1^\dagger\bar{\eta},\cdots,\bar{c}_N^\dagger\bar{\eta})^T$, where $\eta=\text{exp}(i\pi\sum_i n_i)$ is introduced to impose anti-commutation relations between fermionic operators acting from different sides. 
In the first-quantized Liouvillian $\mathbb{L}_0$, the diagonal blocks arise from the effective Hamiltonian, which acts on the density matrix purely from the left or the right, whereas off-diagonal blocks come from quantum jump terms $L_\mu \rho L_\mu^\dagger$, which act on both sides.

\begin{figure}[h]
	\hspace*{-0.03\textwidth}
	\includegraphics[width=1.1\linewidth]{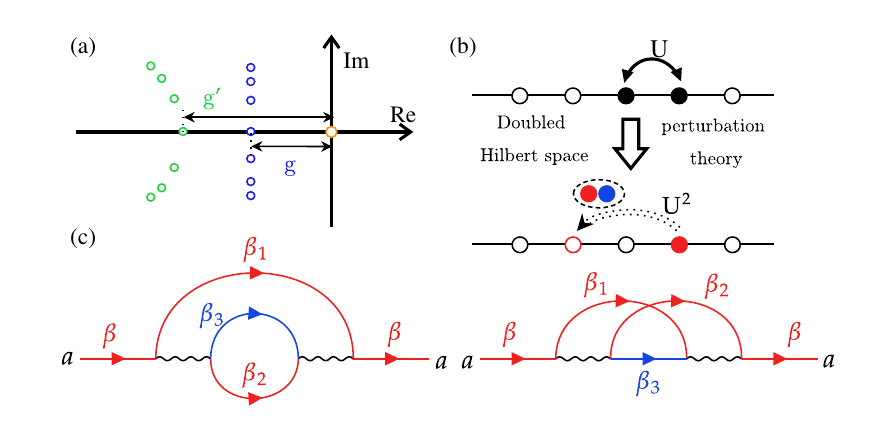}
	\caption{(a) A sketch of Liouvillian spectrum. Blue (green) circles correspond to the model with (without) interactions, and $g$ ($g'$) is the gap. Only eigenvalues close to the imaginary axis are presented. The orange circle corresponds to the steady state. (b) A schematic plot about the self-energy. Interactions between fermions (black dots) with strength $U$ are mapped to the doublon-mediated hopping of a fermionic mode (red dot).  (c) Feynman diagrams of the second-order self-energy. Red (blue) lines correspond to the $a(b)$-fermion, labeled by its complex momentum; wavy lines denote interactions.} 
	\label{fig:illustration}
\end{figure}

Next, we identify bare fermionic modes by block-diagonalizing $\mathbb{L}_0$ in Eq.~\eqref{FreeLind}, which provides the starting point for perturbation expansion. To this end, we employ the non-unitary version of the Bogoliubov transformation:
\begin{equation}
    S=\frac{1}{\sqrt{2}}\left(\begin{array}{cc}
	I & I \\
	I+Z & -I+Z
\end{array} \right),~
\mathbb{L}_0=S^{-1}\left(\begin{array}{cc}
	-X^\dagger & {} \\
	{}& X
\end{array} \right)S.
\end{equation}
Here $X=-ih-(M^g)^T-M^l$, and the matrix $Z$ is chosen to eliminate the off-diagonal blocks by solving the Lyapunov equation $XZ+ZX^\dagger=2(M^g)^T-2M^l$. In the noninteracting setting, these matrices have a direct physical interpretation: $X$ is the damping matrix governing the evolution of correlation functions \cite{Song2019Chiral}, while $Z$ is related to the steady-state correlations by $\lim\limits_{t\to +\infty}\tr{e^{\mathcal{L}_0t}(\rho)c_i^\dagger c_j} = \frac{1}{2}(\delta_{ij}-Z_{ji})$~\cite{Supp}.

Then, we introduce two decoupled sets of fermionic operators as bare modes:
\begin{equation} \label{eq:quasi_particle}
	(\boldsymbol{b}^\dagger,\tilde{\boldsymbol{a}})^T=S\boldsymbol{c},~ (\tilde{\boldsymbol{b}},\boldsymbol{a}^\dagger)=\boldsymbol{c}^\dagger S^{-1},
\end{equation}
in terms of which the non-interacting Liouvillian takes the block-diagonal form: $\mathcal{L}_0=\boldsymbol{a}^\dagger X\tilde{\boldsymbol{a}}-\tilde{\boldsymbol{b}} X^\dagger \boldsymbol{b}^\dagger+\tr{X^\dagger}=\boldsymbol{a}^\dagger X\tilde{\boldsymbol{a}}+\boldsymbol{b}^\dagger X^* \tilde{\boldsymbol{b}}$.
For ease of representation, the main text will focus on the case of balanced gain and loss: $(M^g)^T=M^l$, hence $Z=0$ (infinite-temperature steady state) and $\boldsymbol{a}=\tilde{\boldsymbol{a}}$, $\boldsymbol{b}=\tilde{\boldsymbol{b}}$. 
Our perturbative construction, however, applies to generic $Z$~\cite{Supp}. 
It follows that $X$ ($X^*$) plays the central role of first-quantized non-Hermitian Hamiltonian for bare fermion $\boldsymbol{a}$ ($\boldsymbol{b}$), and the vacuum corresponds to the nonequilibrium steady state of the full Liouvillian.

We are interested in the spectrum of $X$, in particular the eigenvalue with the largest real part (before and after perturbations), which determines the Liouvillian gap [see Fig. \ref{fig:illustration}(a) for a pictorial illustration].
Even in the presence of the Liouvillian skin effect, the Liouvillian gap remains the key quantity governing the long-time asymptotic dynamics, and the boundary relaxation behavior for initial time~\cite{Song2019Chiral, Haga2021Slowing, Yang2022Exactly, Lee2023Anomalously,Yang2025Real,Xue2025Nonbloch}.

\textit{Self-energy and Feynman diagram}---To incorporate density-density interactions $H_I$, we represent the interacting Liouvillian $\mathcal{L}_I=-i\sum_{ i\le j}U_{ij}(n_in_j-\bar{n}_i\bar{n}_j)$ in terms of bare-mode operators~\eqref{eq:quasi_particle}:
\begin{eqnarray}\label{Eqn:Interaction}
    \mathcal{L}_I=-i\sum_{ i\le j}U_{ij}[\frac{1}{4}(a_i^\dagger a_i+a_j^\dagger a_j-b_i^\dagger b_i-b_j^\dagger b_j)\nonumber\\
    +\frac{1}{2}(a_i^\dagger b_i^\dagger+b_i a_i)(a_j^\dagger a_j-b_j^\dagger b_j)+\text{H.c.}],
\end{eqnarray}
which can create or annihilate a pair of $a$- and $b$-fermions conditioned on the local density.
In the case of weak interactions, we adopt the standard perturbation theory to the second order to approximate the pairing potential by the self-energy:
\begin{equation}\label{Eqn:SelfEnergy}
\mathcal{L}_{\text{eff}}(E)=P\mathcal{L}_{0}P+P\mathcal{L}_{I}P+P\mathcal{L}_IQ(E-Q\mathcal{L}_0Q)^{-1}Q\mathcal{L}_IP+\cdots
\end{equation}
Here $P=I-Q$ is the projector of the Hilbert subspace of one $a$-fermion, since the slowest-decaying modes reside in this subspace in non-interacting cases. Therefore, the self-energy is generated as [see Fig.\ref{fig:illustration}(b)]: (i) creating a pair of bare modes (doublon) on the same site, (ii) propagating three bare modes, (iii) annihilating a doublon on another site.

To represent self-energy matrix elements, we introduce single-particle basis states $|i_a\rangle=a_i^\dagger |0\rangle$, and intermediate three-particle states $|i_a,j_a,k_b\rangle$. The zeroth order Liouvillian is given by the damping matrix $\langle i_a|\mathcal{L}_0|j_a\rangle=X_{ij}$ as expected. The first order correction comes from the quadratic term in Eq.~\eqref{Eqn:Interaction}: $\langle i_a|\mathcal{L}_I|j_a\rangle=-\frac{i}{4}\delta_{ij}\sum_k U_{ik}$. 
The first nontrivial contribution appears at second order, which reads:
\begin{align}\label{eq:three_particle}
&\langle i_a|\mathcal{L}_{\text{eff}}^{(2)}(E)|j_a\rangle = \nonumber\\
&-\frac{1}{4}\sum_{kl}U_{ik}U_{jl}\langle i_a,k_a,k_b|(E-Q\mathcal{L}_0Q)^{-1}|j_a,l_a,l_b\rangle.
\end{align}
The central results of this work follow from analyzing this self-energy correction to the damping matrix. We relegate the technical derivations to the Appendices and focus below on the key concepts and essential steps.

We consider lattice-translational invariant damping matrix and interaction strengths: $X_{ij}=X_{i-j}$, $U_{ij}=U_{i-j}$.
Although the translational symmetry is broken under OBC, the self-energy matrix elements in the bulk still respect it approximately. This motivates introducing the \textit{non-Bloch self-energy} $\Sigma^{(2)}(E,\beta)$, which encodes matrix elements into a Laurent polynomial in the complex parameter $\beta$:
\begin{align}\label{Eqn:SelfEnergyGBZ}
\Sigma&^{(2)}(E,\beta)\equiv\sum_{r}\beta^{-r} \langle (i+r)_a|\mathcal{L}_{\text{eff}}^{(2)}|i_a\rangle\nonumber\\
=&-\frac{1}{4}\oint_{\text{GBZ}}\frac{d\beta_1}{2\pi i\beta_1}\oint_{\text{GBZ}}\frac{d\beta_2}{2\pi i\beta_2}\oint_{\text{GBZ}^*}\frac{d\beta_3}{2\pi i\beta_3}\sum_r(\frac{\beta_1\beta_2\beta_3}{\beta})^r \nonumber\\
&\times\frac{U(\beta_2\beta_3)^2-U(\beta_2\beta_3)U(\beta_1\beta_3)}{E-X(\beta_1)-X(\beta_2)-X^* (\beta_3)}.
\end{align}
Here, $X(\beta)=\sum_r \beta^{-r}X_r$, $X^*(\beta)=\sum_r \beta^{-r}X^*_r$, $U(\beta)=\sum_r \beta^{-r}U_{r}$. GBZ${}^*$ is the complex conjugate of GBZ and arises from the damping matrix $X^*$ associated with $b$-fermions.
Two terms in the numerator originate from the indistinguishability of $a$-fermions. This formula sets the starting point for perturbative analysis of Liouvillian eigenvalues and eigenstates.

The appearance of integrals over GBZ is not an artificial choice but a necessary consequence. The derivation involves the calculation of three-particle propagator in Eq.~\eqref{eq:three_particle}, which reduces to Green's functions of non-Hermitian matrix. Under PBC, the Green’s function admits a spectral decomposition in momentum space, i.e., a sum over Bloch modes that becomes an integral over BZ in the thermodynamic limit. By contrast, under OBC, lattice momentum is no longer a good quantum number, and the corresponding spectral representation is instead formulated as an integral over GBZ~\cite{xue2021simple}. 

We further recast Eq.~\eqref{Eqn:SelfEnergyGBZ} in terms of single-particle propagators as demonstrated in End Matter Appendix B, which admits Feynman-diagram representations [Fig.~\ref{fig:illustration}(c)].
There are \textit{external vertices} where a single particle line splits into one particle line and one interaction line. On the other hand, at \textit{internal vertices}, one interaction line transfers momentum to two particle lines, where the conservation law of complex momentum manifests automatically.

We remark that the derivation holds only when the single-particle modes are separated from the intermediate three-particle modes in the real part of eigenvalues. 
Otherwise, interactions would render quasiparticles unstable in the long-time dissipative dynamics, i.e., they are not true eigen modes of the Liouvillian.
This condition manifests as the positivity of the denominator in Eq. \eqref{Eqn:SelfEnergyGBZ} along the integral contour. In Supplemental Material \cite{Supp}, we discuss examples exhibiting instability.

Polynomial coefficients in Eq.~\eqref{Eqn:SelfEnergyGBZ} encode interaction corrections to the damping matrix elements. When it comes to eigenvalues, the spectrum is determined by the self-consistent equation $E(\beta)=E^{(0)}(\beta)+\Sigma^{(2)}(E(\beta),\beta)$, where $E^{(0)}(\beta)=X(\beta),\beta\in\text{GBZ}$ is the non-Bloch band before adding perturbations. For weak interactions, it suffices to evaluate the self-energy on shell: $E(\beta)=E^{(0)}(\beta)+\Sigma^{(2)}(E^{(0)}(\beta),\beta)$.
Under OBC, each eigenmode is composed of $2M$ non-Bloch wave modes ($M$ is the hopping range) with different $\beta$ but the same energy, represented by right and left eigenstates $\ket{R}=\sum_{\mu=1}^{2M}\sum_{i=1}^N\phi_{\mu}^R \beta_\mu^i\ket{i},\bra{L}=\sum_{\mu=1}^{2M}\sum_{i=1}^N\phi_{\mu}^L \beta_\mu^{-i}\bra{i}$.
Here $E^{(0)}(\beta_\mu)=E^{(0)}$ for all $\mu$, where $\beta_\mu$'s are sorted in increasing magnitude \cite{yao2018edge,Yokomizo2019}.
To obtain self-energy corrections to an eigenvalue, one must identify which $\beta$ to take. We find that the relevant correction is given by a weighted average of $\Sigma^{(2)}(E^{(0)},\beta_M)$ and $\Sigma^{(2)}(E^{(0)},\beta_{M+1})$ \cite{Supp}:
\begin{equation}\label{Eqn:Weight}
	\frac{\phi_{M}^R\phi_{M}^L\Sigma^{(2)}(E^{(0)},\beta_M)+\phi_{M+1}^R\phi_{M+1}^L\Sigma^{(2)}(E^{(0)},\beta_{M+1})}{\phi_{M}^R\phi_{M}^L+\phi_{M+1}^R\phi_{M+1}^L}.
\end{equation}
Contributions from other modes are of order $\mathcal{O}(1/N)$ at most.

In practice, evaluation of Eq.~\eqref{Eqn:SelfEnergyGBZ} at fixed $\beta$ is hindered by contributions from large $r$, which obstruct accurate numerical integration. In many cases, we can bypass this difficulty by deforming integral contours from GBZ to BZ, leading to the emergence of complex momentum conservation at \textit{external vertices} which removes the summation over $r$.
Concretely, we deform the $\beta_1$ and $\beta_2$ contours to the BZ and the$\beta_3$ contour to the circle of radius $|\beta|$.
The summation over $r$ then reduces to $\sum_r(\beta_1\beta_2\beta_3/\beta)^r=2\pi i\beta_3\delta(\beta_3-\beta/\beta_1\beta_2)$ by Fourier transform, thereby enforcing momentum conservation. 
See a sufficient condition justifying this deformation in Appendix C, which holds for the model studied in the following sections. It results in the following double integrals on BZ: 
\begin{align}\label{Eqn:SelfEnergyBZ}
\Sigma^{(2)}(&E,\beta)=-\frac{1}{4}\int_0^{2\pi}\frac{dk_1}{2\pi }\int_0^{2\pi}\frac{dk_2}{2\pi }\nonumber\\
&\frac{U(\beta e^{-ik_1})^2-U(\beta e^{-ik_1})U(\beta e^{-ik_2})}{E-X(e^{ik_1})-X(e^{ik_2})-X^* (\beta e^{-ik_1-ik_2})}.
\end{align}

\begin{figure}[h]
	\includegraphics[width=0.46\linewidth]{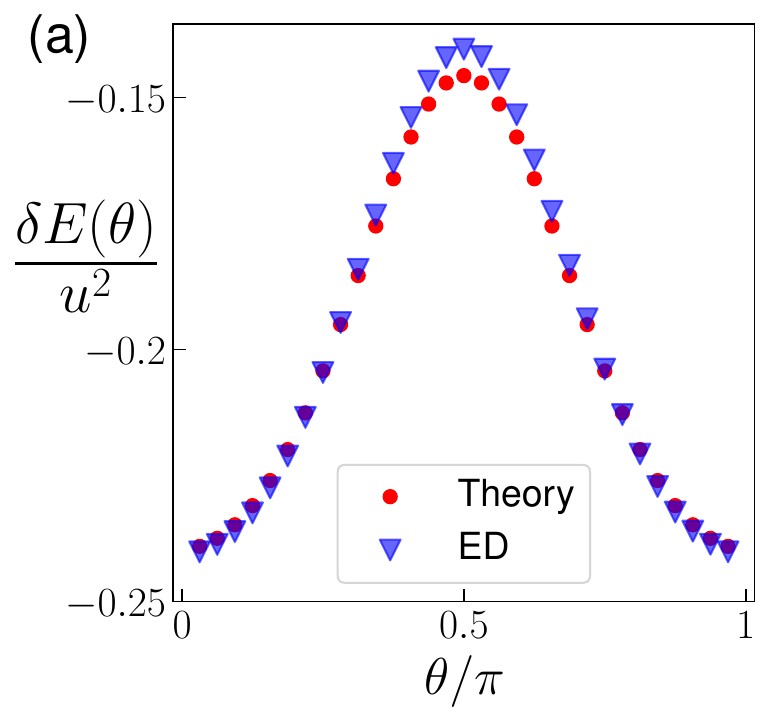}
	\includegraphics[width=0.49\linewidth]{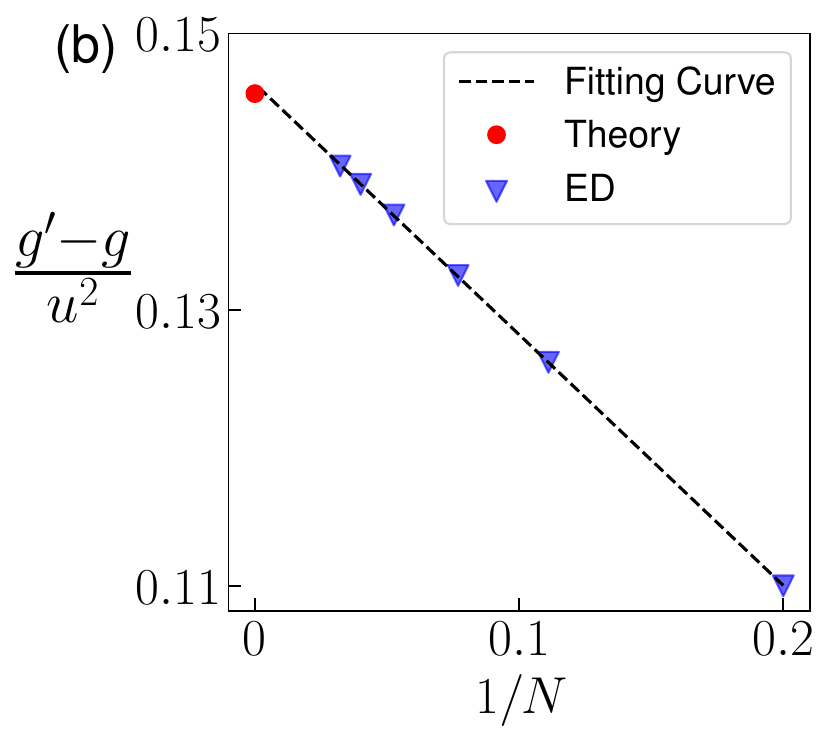}
	\caption{
    Self-energies obtained from Eq.~\eqref{Eqn:SelfEnergyBZ}, compared with exact diagonalization (ED) results. Parameters: $t=1,\gamma=0.5$, interaction strength $u = 0.2$. Corrections are all displayed in the unit of $u^2$.
    (a) Self-energy corrections to eigenvalues $\delta E(\theta)$ as a function of angle $\theta$ on GBZ. 
    Red dots are obtained using Eqs.~(\ref{Eqn:SelfEnergyBZ}, \ref{Eqn:Weight}), and the blue triangles denote ED results. System size $N=31$. 
    (b) Finite-size scaling of the Liouvillian gap correction. The thermodynamic-limit extrapolation approaches the analytical result (the red point). 
    }
	\label{fig:Numerical}
\end{figure} 

\textit{Numerical benchmark}---To benchmark the efficiency and accuracy of our method, we apply the formula to a Liouvillian version \cite{Song2019Chiral} of the Hatano-Nelson model \cite{Hatano1996Localization,Hatano1997Vortex,Hatano1998Non} with nearest-neighbor interactions, in the context of open quantum systems. The corresponding Liouvillian consists of a quadratic Hamiltonian in the first-quantized form $h_{ij}=t(\delta_{i,j+1}+\delta_{i,j-1})$ and quantum jump operators $L_j^l=\sqrt{\frac{\gamma}{2}}(c_j-ic_{j+1}),L_j^g=(L_j^l)^\dagger$. The Laurent polynomial of damping matrix is $X(\beta)=-2\gamma-i(t-\gamma)\beta-i(t+\gamma)\beta^{-1}$, and the interaction term is $U(\beta)=u(\beta+\beta^{-1})$. For $t>\gamma>0$, the GBZ is a circle of radius $R = \sqrt{\frac{t+\gamma}{t-\gamma}}$. 
The spectrum of $a$-fermion is $E^{(0)}(\theta)=-2\gamma-2i\sqrt{t^2-\gamma^2}\text{cos}(\theta)$, where $\theta$ is the angle on GBZ ($\beta = Re^{i\theta}$). 
Since an eigenstate of this damping matrix is a standing wave of two non-Bloch modes with a pair of conjugate complex momentum $R e^{\pm i\theta}$, we use $\theta\in[0, \pi)$ to label eigenstates. 

Before proceeding, we emphasize that even for such a simple NHSE model, the real-space similarity transformation fails to solve the problem. It transforms as $a_j\to R^j a_{j}, a_j^\dagger\to R^{-j} \tilde{a}_{j}^\dagger$ (the same for $b_j$), which attaches the factor $R^{2j}$ to the pairing term $a_jb_j$ [see Eq. \eqref{Eqn:Interaction}] and breaks translational symmetry of the interacting Lindbladian.

We apply Eqs.~(\ref{Eqn:Weight}, \ref{Eqn:SelfEnergyBZ}) to compute self-energy corrections to eigenvalues, and compare them with exact diagonalization (ED) results. As shown in Fig.~\ref{fig:Numerical}(a), the two approaches agree quite well.
Furthermore, in Fig.~\ref{fig:Numerical}(b) we perform finite-size scaling for $\theta=\pi/2$ self-energy, of which the real part is maximal among others. Since the spectrum $E^{(0)}(\theta)$ shares the same real part, the self-energy at $\theta=\pi/2$ determines the Liouvillian gap.
Remarkably, the analytical value lies at the end of the linearly fitting curve of ED. Moreover, the $1/N$ scaling of finite-size corrections agrees with our semi-quantitative approximation of Eq. \eqref{Eqn:Weight}. In Supplemental Material \cite{Supp}, we study a model with non-circular GBZ, and the results support our method further.

\textit{Anisotropic GBZ deformation}---Next, we shift our focus from corrections to eigenvalues to those to eigenstates. To this end, we solve the deformed GBZ, which characterizes the wavefunction localization behavior of perturbed eigenstates. The procedure is as follows: (i) For a given eigenvalue $E^{(0)}$ of the damping matrix $X$ and the associated complex momenta $\beta_M,\beta_{M+1}$, we construct a finite-order Laurent polynomial $E(\beta) = X(\beta) + \Sigma^{(2)}(E^{(0)},\beta)$ by truncating the sum in Eq.~\eqref{Eqn:SelfEnergyGBZ} at an appropriate power $|r|<r_\text{max}$; (ii) Compute the GBZ of $E(\beta)$ and select the two solutions closest to $\beta_M$ and $\beta_{M+1}$; (iii) Repeat it for all $E^{(0)}$. Collecting the selected solutions over the spectrum of $X$ leads to the deformed GBZ.

\begin{figure}[h]
    \hspace*{-0.01\textwidth}
	\includegraphics[width=0.57\linewidth]{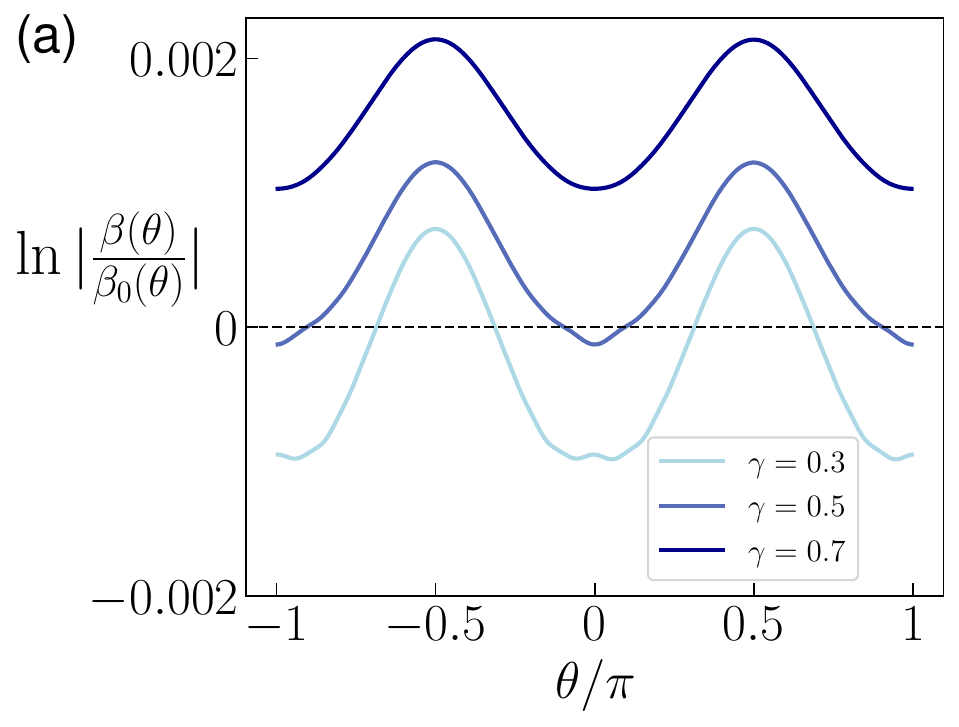}
	\includegraphics[width=0.39\linewidth]{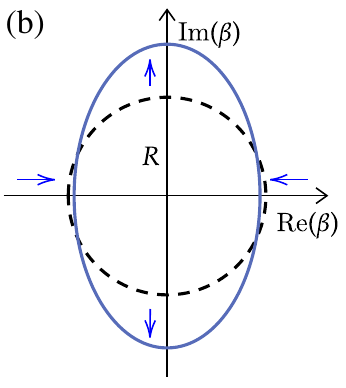}
	\caption{
    Deformation of the generalized Brillouin zone induced by interactions. 
    (a) Angular dependence of the relative modulus change, shown in logarithmic form for several $\gamma$ with fixed parameters $t=1, u=0.2$.
    (b) Illustration of GBZ deformation in the complex plane for $\gamma = 0.5$. Scale is amplified for visibility. The dashed (solid) curve denotes the GBZ before (after) including the self-energy. Blue arrows indicate directions of deformation.
    }
	\label{fig:Squeezing}
\end{figure} 

Fig.~\ref{fig:Squeezing}(a) shows the relative change of the modulus function $|\beta(\theta)|$ along the deformed GBZ for the Liouvillian Hatano–Nelson model at different strengths of non-reciprocity, controlled by $\gamma$.
Remarkably, interactions render the GBZ anisotropic. For weak non-reciprocity ($\gamma=0.3$ and $0.5$), $|\beta(\theta)|$ is suppressed near $\theta=0,\pi$ while being enhanced around $\theta=\pm\frac{\pi}{2}$ [Fig.~\ref{fig:Squeezing}(b)]. For stronger non-reciprocity ($\gamma=0.7$), $|\beta(\theta)|$ increases over the entire angular range. Because the noninteracting GBZ already satisfies $|\beta|>1$, an increase of $|\beta(\theta)|$ signifies a stronger NHSE (i.e., more pronounced skin-mode accumulation), most prominently for the slowest-decaying modes near $\theta=\pm\frac{\pi}{2}$. 

The enhancement of localization can be understood from the schematic plot in Fig.~\ref{fig:illustration}(b). Interactions induce effective hopping by propagating virtual quasiparticles, of which propagators are asymmetric and are stronger along the NHSE localization direction. See Supplemental Material \cite{Supp} for more quantitative analysis and additional parameter sweeps.

Our perturbative framework provides a quasiparticle description of how interactions influence NHSE and GBZ. It should be noted that seemingly contradictory results of NHSE suppression, as reported in Refs.~\cite{mu2020emergent,lee2020many, liu2020non,yoshida2022reduction,alsallom2022fate,dora2022full}, focus on many-body wavefunctions where the Pauli exclusion principle plays a central role. Other works consider the strongly interacting regime where collective modes (e.g., magnons and doublons) can exhibit NHSE~\cite{Brighi2024Nonreciprocal,Yoshida2024Mott}. Both scenarios are essentially distinct from the dissipative quasiparticle picture studied here.

\begin{figure}[h]
	\hspace*{-0.02\textwidth}
	\includegraphics[width=0.48\linewidth]{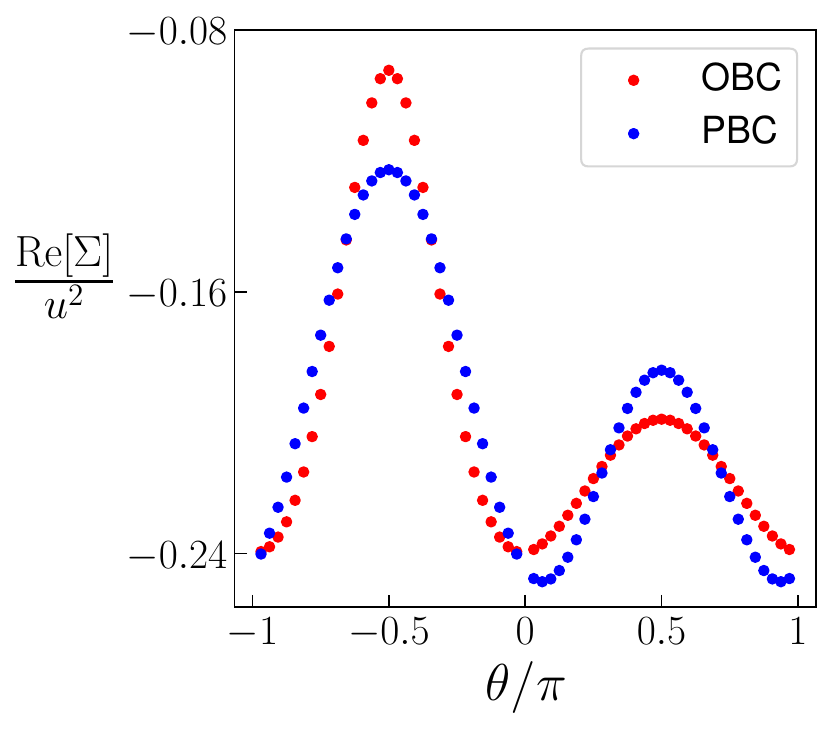}
	\includegraphics[width=0.465\linewidth]{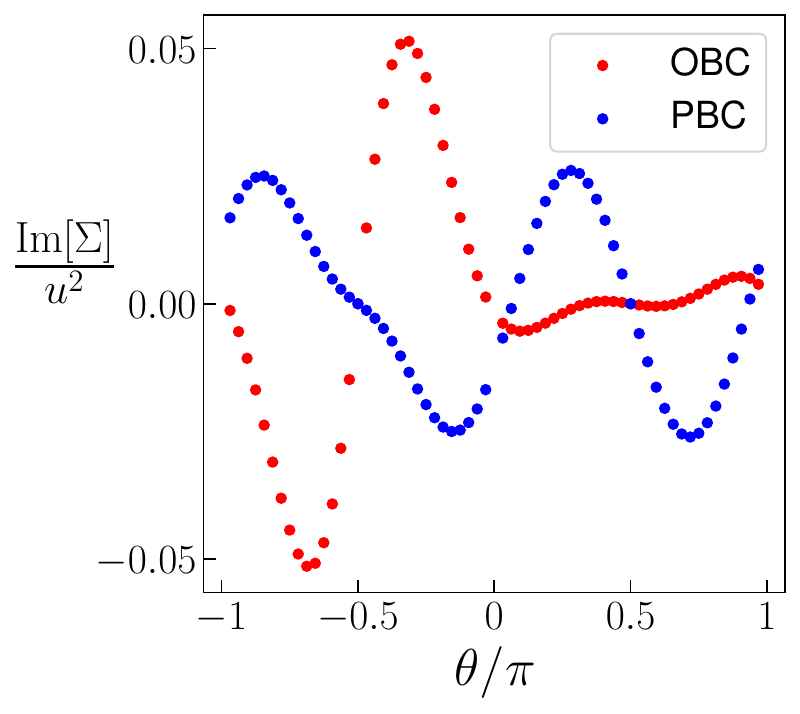}
	\caption{Analytical results of the self-energy for different boundary conditions, as a function of angle of $\beta$ on GBZ (OBC, red dots) and of the momentum (PBC, blue dots). Real and imaginary parts are shown respectively. $t=1,\gamma=0.5$.}
	\label{PBC}
\end{figure}

\textit{Periodic boundary condition---}The self-energy formula can be applied with equal effectiveness to systems under PBC, where the real momentum $k$ becomes a good quantum number. In this case, we should take $\beta=e^{ik}$ in Eq. \eqref{Eqn:SelfEnergyBZ}. In Fig. \ref{PBC}, we compare the self-energy for the same model under different boundary conditions, explicitly showing the boundary sensitivity. Notably, for PBC, the spectrum before perturbations is gapless at $k=-\frac{\pi}{2}$, and interactions open a gap which scales as $u^2$. This is in sharp contrast to the common wisdom for closed 1D systems, where weak interactions are usually non-perturbative and drive the system to the Luttinger liquid phase, characterized by collective low-lying excitations. 
We elucidate the differences through a detailed analysis of the integral formula at the gapless point: to consider the ``low-energy" regime near the singularity $(k_1,k_2)=(-\frac{\pi}{2},-\frac{\pi}{2})$, we introduce the infrared variables $p_{1(2)}=k_{1(2)}+\frac{\pi}{2}$, then the integral is approximated as $\sim \frac{u^2}{\gamma}\int dp_1 dp_2\frac{(p_1^2-p_2^2)-\frac{1}{2}p_1^2(p_1^2-p_2^2)}{p_1^2+p_2^2+p_1p_2}$.
The denominator is of the second order and is symmetric with respect to $p_1$ and $p_2$. In the numerator, the leading asymmetric term vanishes, while the next term is of the fourth order which leads to no divergence.

More generally, the stability of dissipative quasiparticles can be traced to the absence of long-range correlations in the steady state. In a one-dimensional closed fermionic system, perturbation theory fails due to power-law decaying correlations in the ground state~\cite{Giamarchi2003Quantum}. For quadratic fermionic Liouvillians built from local operators, however, steady-state correlations decay exponentially in space~\cite{Zhang2022Criticality,Barthel2022Solving}, thus precluding divergences in perturbative calculations.

\textit{Conclusions}---We develop a systematic perturbation framework for weakly interacting fermionic Markovian open quantum systems exhibiting NHSE.
Our analysis explicitly shows that NHSE leads to complex momentum propagators in the Feynman diagrams when evaluating the self-energy of dissipative quasiparticles.
We present a closed-form integral formula for self-energy corrections to eigenvalues, which agrees well with numerical results.
We also characterize perturbations of Liouvillian eigenstates as deformation of GBZ.
Our study initiates a generic framework to quantitatively analyze the interactions in open quantum systems using non-Bloch band theory, which leads to interaction-dressed dissipative quasiparticles.
Future research could aim at developing the non-Bloch self-energy for multi-band and higher-dimensional systems \cite{Zhang2022Universal,Hu2023Green,Wang2024Amoeba,Zhang2024Algebraic}. 
It would also be interesting to investigate the localization transition of quasiparticles in the strongly interacting regime. 

\textit{Acknowledgments}---We thank Fei Song, Xu-Dong Dai, Dong Yuan, Yu-Min Hu, and Spenser Talkington for helpful discussions. 
This work was supported by the National Natural Science Foundation of China (Grant No. 12125405), and National Key R$\&$D Program of China (No. 2023YFA1406702).

\bibliographystyle{apsrev4-1-title}
\bibliography{Heran_Feynman}

\newpage

\appendix

\makeatletter
\newcommand{\appblock}[1]{%
  \refstepcounter{section}
  \setcounter{equation}{0}
  \renewcommand{\theequation}{\Alph{section}\arabic{equation}}
  \section*{}
}
\makeatother

\section*{End Matter}
\appblock{}
\textit{Appendix A: Self-energy formula on GBZ}---We start from the second-order self-energy in real space [Eq.~\eqref{eq:three_particle}] and derive formulas in the form of integrals over GBZ, with an emphasis on the conditions under which these forms
apply. First, we invoke the GBZ-based Green's function representation in bulk sites~\cite{xue2021simple} to rewrite the three-particle propagator as integrals over GBZ:
\begin{align}
\langle &i_a|\mathcal{L}_{\text{eff}}^{(2)}(E)|j_a\rangle\nonumber\\
=&-\frac{1}{4}\sum_{kl}U_{ik}U_{jl}\langle i_a,k_a,k_b|(E-Q\mathcal{L}_0Q)^{-1}|j_a,l_a,l_b\rangle\nonumber\\
=&-\frac{1}{4}\sum_{kl}U_{ik}U_{jl}\oint_{\text{GBZ}}\frac{d\beta_1}{2\pi i\beta_1}\oint_{\text{GBZ}}\frac{d\beta_2}{2\pi i\beta_2}\oint_{\text{GBZ}^*}\frac{d\beta_3}{2\pi i\beta_3}\nonumber\\
&\frac{(\beta_1^{i-j}\beta_2^{k-l}-\beta_1^{i-l}\beta_2^{k-j})\beta_3^{k-l}}{E-X(\beta_1)-X(\beta_2)-X^* (\beta_3)}.
\end{align}
In the numerator, the term $\beta_1^{i-j}\beta_2^{k-l}$ corresponds to $a$-particles propagating from the site $j$ to $i$ and from $l$ to $k$, and the term $\beta_1^{i-l}\beta_2^{k-j}$ corresponds to propagation from $l$ to $i$ and $j$ to $k$. The relative minus sign originates from fermionic statistics. 

This real-space expression exhibits translational invariance:
\begin{align}
\langle &(i+m)_a|\mathcal{L}_{\text{eff}}^{(2)}(E)|(j+m)_a\rangle\nonumber\\
=&-\frac{1}{4}\sum_{kl}U_{i+m,k}U_{j+m,l}\oint_{\text{GBZ}}\frac{d\beta_1}{2\pi i\beta_1}\oint_{\text{GBZ}}\frac{d\beta_2}{2\pi i\beta_2}\nonumber\\
&\oint_{\text{GBZ}^*}\frac{d\beta_3}{2\pi i\beta_3}\frac{(\beta_1^{i-j}\beta_2^{k-l}-\beta_1^{i-l}\beta_2^{k-j})\beta_3^{k-l}}{E-X(\beta_1)-X(\beta_2)-X^* (\beta_3)}\nonumber\\
=&-\frac{1}{4}\sum_{kl}U_{i+m,k+m}U_{j+m,l+m}\oint_{\text{GBZ}}\frac{d\beta_1}{2\pi i\beta_1}\oint_{\text{GBZ}}\frac{d\beta_2}{2\pi i\beta_2}\nonumber\\
&\oint_{\text{GBZ}^*}\frac{d\beta_3}{2\pi i\beta_3}\frac{(\beta_1^{i-j}\beta_2^{k-l}-\beta_1^{i-l}\beta_2^{k-j})\beta_3^{k-l}}{E-X(\beta_1)-X(\beta_2)-X^* (\beta_3)}\nonumber\\
=&\langle i_a|\mathcal{L}_{\text{eff}}^{(2)}(E)|j_a\rangle,
\end{align}
where we have adopted the translation invariance of interactions: $U_{ij}=U_{i-j}$.
This motivates introducing the Laurent polynomial for the self-energy matrix, as defined in Eq.~\eqref{Eqn:SelfEnergyGBZ}.

\appblock{}
\textit{Appendix B: Feynman-diagrammatic representation}---We rewrite the denominator as a Laplace transform over a real parameter $t$:
\begin{align}
    &\frac{1}{E-X(\beta_1)-X(\beta_2)-X^*(\beta_3)} \nonumber\\
    = &\int_0^\infty dt~e^{-(E-X(\beta_1)-X(\beta_2)-X^* (\beta_3))t},
\end{align}
where $t$ may be interpreted as a time variable. This transformation is only valid when the positivity condition $\Re [E-X(\beta_1)-X(\beta_2)-X^* (\beta_3) ]>0$ holds along the associated integral contours. Physically, this condition amounts to the spectral separation between the single–mode and three-mode sectors, which could be satisfied with a sufficiently large Liouvillian gap. 
For the model studied in the main text, $\Re [E-X(\beta_1)-X(\beta_2)-X^* (\beta_3) ]=4\gamma$, thereby meeting the condition.

With the Laplace transform, the non-Bloch self-energy reads:
\begin{align}\label{eqn:Laplace}
	&\Sigma^{(2)}(E,\beta)=-\frac{1}{4}\oint_{\text{GBZ}}\frac{d\beta_1}{2\pi i\beta_1}\oint_{\text{GBZ}}\frac{d\beta_2}{2\pi i\beta_2}\oint_{\text{GBZ}^*}\frac{d\beta_3}{2\pi i\beta_3} \nonumber\\
	&\sum_r(\frac{\beta_1\beta_2\beta_3}{\beta})^rV(\beta_1,\beta_2,\beta_3)\int_0^\infty dt~e^{-[E-X(\beta_1)-X(\beta_2)-X^* (\beta_3)]t}
\end{align}
Here, we have defined the interaction vertex function $V(\beta_1,\beta_2,\beta_3)=U(\beta_2\beta_3)^2-U(\beta_2\beta_3)U(\beta_1\beta_3)$. 

Then we apply the Fourier transform to rewrite exponential functions as frequency-domain Green's functions:
\begin{equation*}
    e^{X(\beta)t} = \int_0^{2\pi}\frac{d\omega}{2\pi}\frac{e^{i\omega t}}{i\omega-X(\beta)}\equiv\int_0^{2\pi}\frac{d\omega}{2\pi}e^{i\omega t}G_a(i\omega,\beta),
\end{equation*}
\begin{equation}
    e^{X^*(\beta)t} = \int_0^{2\pi}\frac{d\omega}{2\pi}\frac{e^{i\omega t}}{i\omega-X^*(\beta)}\equiv\int_0^{2\pi}\frac{d\omega}{2\pi}e^{i\omega t}G_b(i\omega,\beta).
\end{equation}
The Fourier transform holds when $\Re [X(\beta)]\le 0$ for $\beta$ on GBZ. In our setting, this condition is automatically met, since $X(\beta)$ are eigenvalues of the single-particle Liouvillian
and therefore have non-positive real parts.

With the Fourier transform, the non-Bloch self-energy reads
\begin{align}
	&\Sigma^{(2)}(E,\beta)=-\frac{1}{4}\oint_{\text{GBZ}}\frac{d\beta_1}{2\pi i\beta_1}\oint_{\text{GBZ}}\frac{d\beta_2}{2\pi i\beta_2}\oint_{\text{GBZ}^*}\frac{d\beta_3}{2\pi i\beta_3} \nonumber\\
    &\sum_r(\frac{\beta_1\beta_2\beta_3}{\beta})^r
	\int_0^\infty dt\int_0^{2\pi}(\frac{d\omega_1}{2\pi}\frac{d\omega_2}{2\pi}\frac{d\omega_3}{2\pi})e^{(i\omega_1+i\omega_2+i\omega_3-E)t}\nonumber\\
	&\times V(\beta_1,\beta_2,\beta_3)G_a(i\omega_1,\beta_1)G_a(i\omega_2,\beta_2)G_b(i\omega_3,\beta_3).
\end{align}
This expression corresponds to Feynman diagrams shown in Fig.~\ref{fig:illustration}(c), where red and blue lines represent Green's functions $G_a(i\omega,\beta)$ and $G_b(i\omega,\beta)$, respectively.

\appblock{}
\textit{Appendix C: Self-energy formula on BZ}---Next, we deform integral contours from GBZ to BZ to simplify the form. Since now the integrand in Eq.~\eqref{eqn:Laplace} exhibits no poles other than the original point, we can deform the integral contours of $\beta_{1,2}$ to the Brillouin zone through the parameterization $\beta_{1,2} =e^{ik_{1,2}}$. Accordingly, the integral contour of $\beta_3$ is deformed to the circle of radius $|\beta|$, with parameterization $\beta_3 = \beta e^{ik_3}$. Consequently, the sum over $r$ reads
\begin{equation}
    \sum_r (\frac{\beta_1\beta_2\beta_3}{\beta})^r = \sum_r e^{i(k_1+k_2+k_3)r}=2\pi \delta(k_1+k_2+k_3).
\end{equation}
It follows that the self-energy formula is reduced to a double-integral over BZ:
\begin{widetext}
\begin{align}
\Sigma^{(2)}(E,\beta)=&-\frac{1}{4}\int_0^{2\pi}\frac{dk_1}{2\pi }\int_0^{2\pi}\frac{dk_2}{2\pi }\int_0^{2\pi}\frac{dk_3}{2\pi }
 2\pi \delta(k_1+k_2+k_3)V(e^{ik_1},e^{ik_2},\beta e^{ik_3}) \int_0^\infty dt e^{-[E-X(e^{ik_1})-X(e^{ik_2})-X^* (\beta e^{ik_3})]t}\nonumber\\
=&-\frac{1}{4}\int_0^{2\pi}\frac{dk_1}{2\pi }\int_0^{2\pi}\frac{dk_2}{2\pi } V(e^{ik_1},e^{ik_2},\beta e^{-ik_1-ik_2}) \int_0^\infty dt e^{-[E-X(e^{ik_1})-X(e^{ik_2})-X^* (\beta e^{-ik_1-ik_2})]t}\nonumber\\
=&-\frac{1}{4}\int_0^{2\pi}\frac{dk_1}{2\pi }\int_0^{2\pi}\frac{dk_2}{2\pi }\frac{V(e^{ik_1},e^{ik_2},\beta e^{-ik_1-ik_2})}{E-X(e^{ik_1})-X(e^{ik_2})-X^* (\beta e^{-ik_1-ik_2})}.
\end{align}
\end{widetext}

For the last equality, we have performed the integral over the real parameter $t$, which is allowed only when
\begin{equation}
    \Re[E-X(e^{ik_1})-X(e^{ik_2})-X^* (\beta e^{-ik_1-ik_2})]>0
\end{equation}
for $k_{1,2}$ on BZ. This condition should be distinguished from the positivity requirement of the denominator in Eq.~\eqref{Eqn:SelfEnergyGBZ}: the latter encodes the physical stability of quasiparticles in the presence of interactions, whereas the former is a purely mathematical condition required to deform the integral contour from GBZ to BZ.

\end{document}


\title{Supplementary Material for: Non-Bloch self-energy of dissipative interacting fermions}
\author{He-Ran Wang}
\thanks{These authors contributed equally to this work.}
\affiliation{Institute for Advanced Study, Tsinghua University, Beijing 100084,
People's Republic of China}
\author{Zijian Wang}
\thanks{These authors contributed equally to this work.}
\affiliation{Institute for Advanced Study, Tsinghua University, Beijing 100084,
People's Republic of China}
\author{Zhong Wang}
\email{wangzhongemail@tsinghua.edu.cn}

\affiliation{Institute for Advanced Study, Tsinghua University, Beijing 100084,
People's Republic of China}

\maketitle

\section{The framework of free-fermion Liouvillian}\label{sec:Frame}
\subsection{Fermionic doubled Hilbert space}
In this subsection, we will work out the formulation of the free-fermionic Liouvillian in detail. To map the fermionic Liouvillian to a rank-2 non-Hermitian Hamiltonian, we apply the technique of doubled Hilbert space. The mapping of action of operators on the Fock basis is given by:
\begin{align}
&|m\rangle\langle n|\to|m\rangle\otimes |n\rangle,~c_i^\dagger |m\rangle\langle n|\to c_{i}^\dagger |m\rangle\otimes |n\rangle,~c_i |m\rangle\langle n|\to c_{i}|m\rangle\otimes |n\rangle,\nonumber\\
&|m\rangle\langle n|c_i\to\bar{c}_{i}^\dagger |m\rangle\otimes |n\rangle,  |m\rangle\langle n|c_i^\dagger\to\bar{c}_{i} |m\rangle\otimes |n\rangle.
\end{align}
However, under such mapping, fermionic operators on different sides of the density matrix $c_i$ and $\bar{c}_j$ commute with each other, $[c_i,\bar{c}_j]=0$. 
To impose mutual anti-commutation relations, we introduce the Klein factor multiplied $\bar{c}$ and $\bar{c}^\dagger$ operators as in the literature of bosonization \cite{Von1998Bosonization}: $\bar{c}_i\to K\bar{c}_i, K=\prod_i (-1)^{n_i+\bar{n}_i}$. Then, we obtain the desired mutual statistics:
\begin{equation}
    \{c_i,\bar{c}_j\}=c_i\bar{c}_j+\bar{c}_jc_i=-Kc_iK\bar{c}_j+KK\bar{c}_jc_i=-K[c_i,K\bar{c}_j]=0.
\end{equation}
It follows that $K=\eta\bar{\eta}$, 
where $\eta=\prod_i (-1)^{n_i}[\bar{\eta}=\prod_i (-1)^{\bar{n}_i}]$ is the fermionic parity operator for left (right) fermions. Following the mapping rule, we obtain the double-space representation of the quadratic Liouvillian as the main text:
\begin{equation}
     \mathcal{L}_0=\mathbf{c}^\dagger\mathbb{L}_0\mathbf{c}-\tr{M^l+(M^g)^T-ih},
\qquad\mathbb{L}_0=\left(\begin{array}{cc}
	-ih+(M^g)^T-M^l & 2(M^g)^T \\
	2M^l& -ih-(M^g)^T+M^l
\end{array} \right).
 \end{equation}
The $2N\times 2N$ matrix $\mathbb{L}_0$, as the first quantized Liouvillian, can be block diagonalized by a similarity transformation.
We will then demonstrate how to construct the transformation matrix stepwisely. 

First, we apply the unitary transformation $H=\frac{1}{\sqrt{2}}\left( \begin{array}{cc}
I & I \\
I & -I
\end{array} \right)$ to block-lower-diagonalize $\mathbb{L}_0$:
\begin{eqnarray}
H\mathbb{L}_0 H^{-1}=\left( \begin{array}{cc}
-X^{\dagger} & 0 \\
2[(M^g)^T-M^l] &X
\end{array} \right).
\end{eqnarray}
This transformation corresponds to the Keldysh rotation in the context of nonequilibrium field theory~\cite{Kamenev2011Field}.
Next, following the standard method for diagonalizing lower-triangle matrices, we introduce the transformation $T=\left( \begin{array}{cc}
I & 0 \\
Z & I
\end{array} \right)$ that acts as
\begin{equation}
\left( \begin{array}{cc}
I & 0 \\
Z & I
\end{array} \right)
\left( \begin{array}{cc}
-X^{\dagger} & 0 \\
2[(M^g)^T-M^l] &X
\end{array} \right)
\left( \begin{array}{cc}
I & 0 \\
-Z & I
\end{array} \right)=
\left( \begin{array}{cc}
-X^{\dagger} & 0 \\
2[(M^g)^T-M^l]-XZ-ZX^\dagger &X
\end{array} \right).
\end{equation}
The matrix $Z$ is determined by the condition that the off-diagonal block vanishes:
\begin{equation}\label{eq:Lyapunov}
    2[(M^g)^T-M^l]-XZ-ZX^\dagger=0,
\end{equation}
which admits the form of the Lyapunov equation. Consequently, we obtain the desired similarity transformation $S$ which block diagonalizes $\mathbb{L}_0$:
\begin{equation}
    S=TH=\frac{1}{\sqrt{2}}\left(\begin{array}{cc}
	I & I \\
	I+Z & -I+Z
\end{array} \right), \qquad S\mathbb{L}_0S^{-1}=\left(\begin{array}{cc}
	-X^\dagger & {} \\
	{}& X
\end{array} \right).
\end{equation}

\subsection{The non-equilibrium steady state}
Real part of the damping matrix spectrum is semi-definite negativity, guranteed by the dissipative nature of the Liouvillian dynamics.
The non-equilibrium steady state $|\rho_\text{ss}\rangle$ , being the zero-eigenvalue eigenstate of the Liouvillian, is the vacuum of $a-$ and $b-$fermions:
\begin{equation}\label{eqn:unocc}
\tilde{a}_{m}|\rho_\text{ss}\rangle=\tilde{b}_{m}|\rho_\text{ss}\rangle=0,~\forall m.
\end{equation}
For the quadratic Liouvillian, we propose the fermionic Gaussian state ansatz for the steady state (without normalization):
\begin{equation}
	\rho_\text{ss}=\text{exp}[\sum_{mn}(G)_{mn}c_m^\dagger c_n],\qquad|\rho_\text{ss}\rangle=\text{exp}[\sum_{mn}(e^G)_{mn}c_m^\dagger \bar{c}_n^\dagger]|0,0\rangle,
\end{equation}
where $G$ is a Hermitian matrix to be decided, dubbed as the first quantized modular Hamiltonian. We diagonalize $G$ as $G_{ij}=\sum_{k} V^*_{ki}V_{kj}g_k$ and define fermion operators $d_k=\sum_l V_{kl}c_l,~\bar{d}_k=\sum_l V^*_{kl}\bar{c}_l$, then the steady state admits the canonical form:
\begin{equation}
	|\rho_\text{ss}\rangle=\text{exp}(\sum_{k}e^{g_k}d_k^\dagger \bar{d}_k^\dagger)|0,0\rangle=\prod_k[(1+e^{g_k})d_k^\dagger \bar{d}_k^\dagger]|0,0\rangle.
\end{equation}
The state fulfills relations
\begin{equation}\label{eqn:condition}
	(1+e^{g_k})\eta d_k|\rho_\text{ss}\rangle=(1-e^{g_k})\bar{d}_k^\dagger\bar{\eta}|\rho_\text{ss}\rangle,\qquad
	(1-e^{g_k}) d_k^\dagger\eta|\rho_\text{ss}\rangle=-(1+e^{g_k})\bar{\eta}\bar{d}_k|\rho_\text{ss}\rangle,
\end{equation}

On the other hand, we can rewrite the annihilation condition \eqref{eqn:unocc} by diagonalizing the matrix $Z$ as $Z_{ij}=\sum_k W^*_{ki}W_{kj} z_k$, and define fermions:$f_k=\sum_l W_{kl}c_l,~\bar{f}_k=\sum_l W^*_{kl}\bar{c}_l$, then we have
\begin{equation}
(1+z_k)\eta f_k|\rho_\text{ss}\rangle=(1-z_k)\bar{f}_k^\dagger\bar{\eta}|\rho_\text{ss}\rangle,\qquad
(1-z_k) f_k^\dagger\eta|\rho_\text{ss}\rangle=-(1+z_k)\bar{\eta}\bar{f}_k|\rho_\text{ss}\rangle.
\end{equation}
Compare it with Eq. \eqref{eqn:condition}, we can explicitly establish the relation between $G$ and $Z$:
\begin{equation}
	G=\ln[\frac{I-Z}{I+Z}].
\end{equation}
The normalization factor (partition function) of $\rho_\text{ss}$ is given by
\begin{equation}
\mathcal{Z}=\text{Tr}(\rho_\text{ss})=\prod_k (1+\frac{1-z_k}{1+z_k})=\frac{2^N}{\text{Det}(1+Z)}.
\end{equation}
In the main text, we consider the $Z=0$ case, where the steady state is the infinite-temperature state, in accordance with the common knowledge that the Liouvillian dynamics with Hermitian dissipators (notice that dissipators are defined up to a unitary transformation over different channels) induces purely decoherence and leads to the maximally mixed state.

\subsection{$Z$-matrix for lattice models}
The above analysis shows that the spectrum of quadratic fermion Liouvillian fully depends on the damping matrix $X$, while the $Z$-matrix characterizes the Gaussian steady state. In this subsection. we will discuss the explicit form of the $Z$-matrix for lattice models under the open boundary condition, through solving the Lyapunov equation~\eqref{eq:Lyapunov}. Then, we will derive the self-energy formula for a generic $Z$-matrix, as a complement to $Z = 0$ results in the main text.

We unfold the matrix $Z$ to a vector $\ket{Z}=\sum_{ij}Z_{ij}\ket{i}\otimes\ket{j}$, where the Lyapunov equation is subsequently transformed to $(X\otimes I+I\otimes X^*)|Z\rangle=|M\rangle$,  $M=2((M^g)^T-M^l)$. Then we tend to take the inverse of $X\otimes I+I\otimes X^\dagger$ to evaluate $Z$. To this end, we perform spectral decomposition for the inverse matrix as:
\begin{eqnarray}
[\frac{1}{X\otimes I+I\otimes X^*}]_{ij,lk}&=&\sum_{mn}\frac{1}{\lambda_m+\lambda_n^*}\langle i|u_{Rm}\rangle\langle u_{Lm}|j\rangle\langle k|u_{Ln}\rangle\langle u_{Rn}|l\rangle\nonumber\\
&=&\sum_m \langle i|u_{Rm}\rangle\langle u_{Lm}|j\rangle\sum_n \frac{\langle k|u_{Ln}\rangle\langle u_{Rn}|l\rangle}{\lambda_m+\lambda_n^*}\nonumber\\
&=&-\sum_m \langle i|u_{Rm}\rangle\langle u_{Lm}|j\rangle G_{lk}(-\lambda_m,X^*).
\end{eqnarray}
Here, the right and left eigenstates are defined as $X|u_{Rm}\rangle=\lambda_m|u_{Rm}\rangle,X^\dagger|u_{Lm}\rangle=\lambda_m^*|u_{Lm}\rangle$, and the single-particle Green's function is $G_{ij}(z,X)=[\frac{1}{z-X}]_{ij}$. Before proceeding, we introduce two formulas. The first one comes from the complex analysis:
\begin{equation}
\frac{\partial}{\partial z^*}\frac{1}{z}=\pi\delta(x)\delta(y)=\pi\delta(z),
\end{equation}
here $z=x+iy$ is defined on the complex plane, and $\frac{\partial}{\partial z^*}=\frac{1}{2}(\frac{\partial}{\partial x}+i\frac{\partial}{\partial y})$. The other one is the formula of Green's function under open boundary conditions~\cite{xue2021simple} :
\begin{equation}
G_{ij}(z,X)=\oint_{\text{GBZ}}\frac{d\beta}{2\pi i\beta}\frac{\beta^{i-j}}{z-X(\beta)}.
\end{equation}
In general, one should conduct the integral on a convergence circle, on which the complex function $z-X(\beta)$ has a vanishing winding number with respect to the zero point. Such an integral contour can always be replaced by the GBZ, since the image of $X(\beta)$ from the GBZ is indeed the spectrum of $X$ under open boundary conditions, which winds zero with respect to any point.

We calculate the matrix inverse utilizing the two formulas:
\begin{eqnarray}
[\frac{1}{X\otimes I+I\otimes X^*}]_{ij,lk}&=&-\int d^2 z [\delta(z-X)]_{ij}G_{lk}(-z,X^*)=-\int d^2z [\frac{\partial}{\pi\partial z^*} G_{ij}(z,X)]G_{lk}(-z,X^*)\nonumber\\
&=&-\int d^2z [\frac{\partial}{\pi\partial z^*}\oint_{\text{GBZ}}\frac{d\beta}{2\pi i\beta}\frac{\beta^{i-j}}{z-X(\beta)}]G_{lk}(-z,X^*)\nonumber\\
&=&-\int d^2z \oint_{\text{GBZ}}\frac{d\beta}{2\pi i\beta} \beta^{i-j}\delta(z-X(\beta))G_{lk}(-z,X^*)\nonumber\\
&=&\oint_{\text{GBZ}}\frac{d\beta}{2\pi i\beta}\oint_{\text{GBZ}^*}\frac{d\beta'}{2\pi i\beta'}\frac{\beta^{i-j}{\beta'}^{l-k}}{X(\beta)+X^*(\beta')}\nonumber\\
&=&\int_{-\infty}^{+\infty}\frac{d\omega}{2\pi}\oint_{\text{GBZ}}\frac{d\beta}{2\pi i\beta}\frac{\beta^{i-j}}{i\omega+X(\beta)}\oint_{\text{GBZ}^*}\frac{d\beta'}{2\pi i\beta'}\frac{{\beta'}^{l-k}}{i\omega-X^*(\beta')}
\end{eqnarray}

At first sight, the matrix inverse depends on the boundary condition. However, for fermionic systems, the real part of $X(\beta)$ on the GBZ or BZ can never be positive due to the dissipative nature of the Liouvillian under both open and periodic boundary conditions. Therefore, we can freely deform the integral contour between the two. At least for $i,j$ deep in the bulk, the corresponding matrix elements of $Z$ should be the same for PBC and OBC: $Z_{ij}=\int_0^{2\pi} \frac{dk}{2\pi} Z(k)e^{ik(i-j)}$.

In the main text, we have assumed $(M^g)^T= M^l$ for simplicity. Here, for the generic case $(M^g)^T\neq M^l$, we will repeat the derivation of self-energy. The interacting Liouvillian $\mathcal{L}_I$ reads
\begin{equation}
    \mathcal{L}_I = -i\sum_{\langle i j\rangle}U_{ij}(n_in_j - \bar{n}_i\bar{n}_j) = -i\sum_{\langle i j\rangle}U_{ij}(n_i - \bar{n}_i)(n_j + \bar{n}_j).
\end{equation}
It can be represented in terms of bare modes as
\begin{equation}
    n_i-\bar{n}_i=a_{i}^\dagger \tilde{a}_{i}-b_{i}^\dagger \tilde{b}_{i},
    \end{equation}
and
    \begin{equation}
     n_i+\bar{n}_i=1-Z_{ii}+\sum_j (Z_{ij}b_{j}^\dagger \tilde{b}_{i}+Z_{ji}a_{j}^\dagger \tilde{a}_{i})
     +\tilde{b}_{i}\tilde{a}_{i}+a_{i}^\dagger b_{i}^\dagger-\sum_{jk}Z_{ij}Z_{ik}a_{j}^\dagger b_{k}^\dagger,
\end{equation}
such that the second-order perturbation term in the real space is given by
\begin{align}
\langle i_a|\mathcal{L}_{\text{eff}}^{(2)}|j_a\rangle=-\frac{1}{4}(\sum_{kl}U_{ik}U_{jl}-\sum_{knlm}U_{ik}U_{jn}Z_{nl}Z_{nm})\oint_{\text{GBZ}}\frac{d\beta_1}{2\pi i\beta_1}\oint_{\text{GBZ}}\frac{d\beta_2}{2\pi i\beta_2}\oint_{\text{GBZ}}\frac{d\beta_3}{2\pi i\beta_3}
\frac{(\beta_1^{i-j}\beta_2^{k-l}-\beta_1^{i-l}\beta_2^{k-j})\beta_3^{k-l}}{E-X(\beta_1)-X(\beta_2)-X^* (\beta_3)}.
\end{align}
By imposing the translational invariant, we can calculate the non-Bloch self-energy as the Laurent polynomial of the self-energy matrix. As a result, an additional prefactor $1-Z(\beta_2)Z(\beta_3)$ is attached to the integrand:
\begin{equation}
\Sigma^{(2)}(E,\beta)
=-\frac{1}{4}\oint_{\text{GBZ}}\frac{d\beta_1}{2\pi i\beta_1}\oint_{\text{GBZ}}\frac{d\beta_2}{2\pi i\beta_2}\oint_{\text{GBZ}^*}\frac{d\beta_3}{2\pi i\beta_3}\sum_r(\frac{\beta_1\beta_2\beta_3}{\beta})^r \times[1-Z(\beta_2)Z(\beta_3)]\frac{U(\beta_2\beta_3)^2-U(\beta_2\beta_3)U(\beta_1\beta_3)}{E-X(\beta_1)-X(\beta_2)-X^* (\beta_3)}.
\end{equation}

\section{Stability of quasiparticles}

In the main text, we have demonstrated the spectral separation between single-particle and three-particle sectors as a requirement for perturbative analysis. Physically, this requirement ensures the stability of quasiparticles against decay into multi-particle states.

For the model considered in the main text, with $X(\beta) = -2\gamma-i(t-\gamma)\beta- 2i(t+\gamma)\beta^{-1}$, the spectrum takes different forms depending on the relative magnitude of $t$ and $\gamma$. When $t>\gamma>0$, it reads $E^{(0)}(\theta) = -2\gamma - i\sqrt{t^2-\gamma^2}\cos(\theta), \theta \in [0,\pi)$. In this regime, all single-particle eigenvalues share the same real part, so the two sectors are separated by $4\gamma$.

On the other hand, when $\gamma>t>0$, the spectrum becomes purely real: $E^{(0)}(\theta) = -2\gamma - 2\sqrt{\gamma^2-t^2}\cos(\theta), \theta \in [0,\pi)$. Single-particle eigenvalues then occupy the interval $(-2\gamma - 2\sqrt{\gamma^2-t^2}, -2\gamma + 2\sqrt{\gamma^2-t^2})$, while the three-particle spectrum spans $(-6\gamma - 6\sqrt{\gamma^2-t^2}, -6\gamma + 6\sqrt{\gamma^2-t^2})$. These intervals overlap when $\gamma^2<3t^2/4$, in which case the perturbative treatment breaks down and the self-energy calculation would meet divergence.\\

\section{Weighted average of the non-Bloch self-energy}
\subsection{NHSE eigenstates and the boundary condition}
For 1D open boundary lattice models without disorders, the generic single-band non-Hermitian Hamiltonian takes the form $H=\sum_i\sum_{l=-M}^M t_{l}\ket{i}\bra{i+l}$, where $M$ is the hopping range. For simplicity, here we focus on the symmetric hooping range. We take an ansatz for the (right) eigenstate of $H$ with eigenvalue $E$ as $|R\rangle=\sum_{\mu=1}^{2M}\sum_{i=1}^N\phi_\mu^R\beta_\mu^i|i\rangle=\sum_{i=1}^N \psi_i\ket{i}$, where $\beta_{\mu}$ are 2$M$ solutions of the characteristic equation $E=H(\beta)=\sum_{l=-M}^M t_l \beta^{-l}$, ordered by their modulus. The coefficients $\phi^R_\mu$ are fixed by imposing the open boundary condition such that $\psi_{1-\mu}=\psi_{N+\mu}=0,\mu=1,\cdots,M$. Thus We obtain 2$M$ equations of $\phi_\mu^R$ as
\begin{equation}
	\sum_{\mu=1}^{2M}\beta_{\mu}^{1-\nu} \phi_\mu^R=0,\quad\sum_{\mu=1}^{2M}\beta_{\mu}^{N+\nu}\phi_\mu^R=0,\quad \nu=1,\cdots,M.
\end{equation}
Linear equations of $\phi_\mu^L$ can also be derived following the same method. 

According to the non-Bloch band theory \cite{yao2018edge,Yokomizo2019}, for an eigen-energy $E$, among the roots of the equation $H(\beta)=E$, we have $|\beta_M|=|\beta_{M+1}|$. If we choose $\phi_\mu^R,\phi_\mu^L$ to be $\mathcal{O}(1)$ when $\mu=M,M+1$, fulfilling the boundary condition requires that $\phi_\mu^{R(L)}\sim\mathcal{O}(1)$ for $\mu<M(\mu>M+1)$ and $\phi_\mu^{R(L)}\sim\mathcal{O}(|\frac{\beta_M}{\beta_\mu}|^N)(\mathcal{O}(|\frac{\beta_\mu}{\beta_M}|^N))$ for $\mu>M+1(\mu<M)$. Notice that with disordered terms on the boundary, the explicit form of the boundary equations will be changed, though the above conclusions about the scaling behavior of the coefficients are still preserved.

\subsection{Scaling analysis for the self-energy}
By definition, the second-order energy shift for an eigenstate $|R\rangle$ of the damping matrix is $\langle L|\hat{\Sigma}^{(2)}(E)|R\rangle/\langle L|R\rangle$. Exploiting the translational invariance of the self-energy matrix elements $\Sigma^{(2)}_{ij}$, we can express the expectation value of self-energy as a summation of $\Sigma^{(2)}(\beta_\mu,E)$ over $\mu=1$ to $2M$, with different weights:
\begin{equation}\label{Eqn:order}
    \langle L|\hat{\Sigma}^{(2)}(E)|R\rangle=\sum_{\mu\nu}^{2M}[\phi_\mu^R\phi_\nu^L\Sigma^{(2)}(\beta_\mu,E)\sum_{j=1}^N (\beta_\mu/\beta_\nu)^j]\equiv\sum_{\mu\nu}^{2M} \sigma_{\mu,\nu}.
\end{equation}
We will prove that in the thermodynamic limit, among all pairs of index $\mu,\nu$, two of them $\sigma_{M,M}, \sigma_{M+1,M+1}$ dominate the others. 

Given the relative magnitude of different $\beta_\mu$'s in the above subsection, the scaling behavior of terms in Eq.~\eqref{Eqn:order} can be categorized into five scenarios:
\begin{itemize}
    \item $\mu=\nu=M$, $\sigma_{M,M}=N\phi_M^R\phi_M^L\Sigma^{(2)}(\beta_M,E)$. Similarly, $\sigma_{M+1,M+1}=N\phi_{M+1}^R\phi_{M+1}^L\Sigma^{(2)}(\beta_{M+1},E)$, and the two terms are of the same order $\sim \mathcal{O}(N)$.
    \item $\mu=M,\nu=M+1$ or $\mu=M+1,\nu=M$. Define the phase factor $e^{i\phi}=\beta_{M+1}/\beta_M$, in $\sigma_{\mu,\nu}$ the summation $\sum_{j=1}^N (e^{\pm i\phi})^j$ goes to zero in the thermodynamic limit, since $\beta_M\neq\beta_{M+1}$ in general.
    \item $\mu=\nu\neq M,M+1$. If $\mu>M+1$, then $\sigma_{\mu,\mu}\sim N|\beta_M/\beta_\mu|^N$, otherwise $\mu<M$ and $\sigma_{\mu,\mu}\sim N|\beta_\mu/\beta_M|^N$, both is much smaller than the first case.
    \item $\mu>\nu$ and $ |\beta_\mu|\neq|\beta_\nu|$. If $\mu\le M+1$, we have $\phi_\mu^R\sim \mathcal{O}(1),\phi_\nu^L\sim |\beta_\nu/\beta_M|^N$, so $\sigma_{\mu,\nu}\sim|\beta_\mu/\beta_M|^N$. If $\mu> M+1$ and $\nu< M$, $\sigma_{\mu,\nu}\sim\mathcal{O}(1)$. Otherwise, $\nu\ge M$ and $\sigma_{\mu,\nu}\sim|\beta_M/\beta_\nu|^N$. All of them are much smaller than $\mathcal{O}(N)$.
    \item $\mu<\nu$ and $ |\beta_\mu|\neq|\beta_\nu|$. The analysis is very similar with the previous one, so as the conclusion.
\end{itemize}
The above qualitative analysis shows that leading contributions to the self-energy scale as $\mathcal{O}(N)$ (without eigenstate normalization), from $\sigma_{M,M}+\sigma_{M+1,M+1}$, while other terms scale as $\mathcal{O}(1)$ at most. Therefore, taking the normalization into account, the error of our formula scales as $\mathcal{O}(1/N)$. Adding the normalization factor, a closed-form eigenstate self-energy is given by:
\begin{equation}
    \langle L|\hat{\Sigma}^{(2)}(E)|R\rangle\simeq (\phi_M^R\phi_M^L\Sigma^{(2)}(\beta_M,E)+\phi_{M+1}^R\phi_{M+1}^L\Sigma^{(2)}(\beta_{M+1},E))/(\phi_M^R\phi_M^L+\phi_{M+1}^R\phi_{M+1}^L).
\end{equation}

\begin{figure}[h]
	\hspace*{-0.36\textwidth}
	\includegraphics[width=0.28\linewidth]{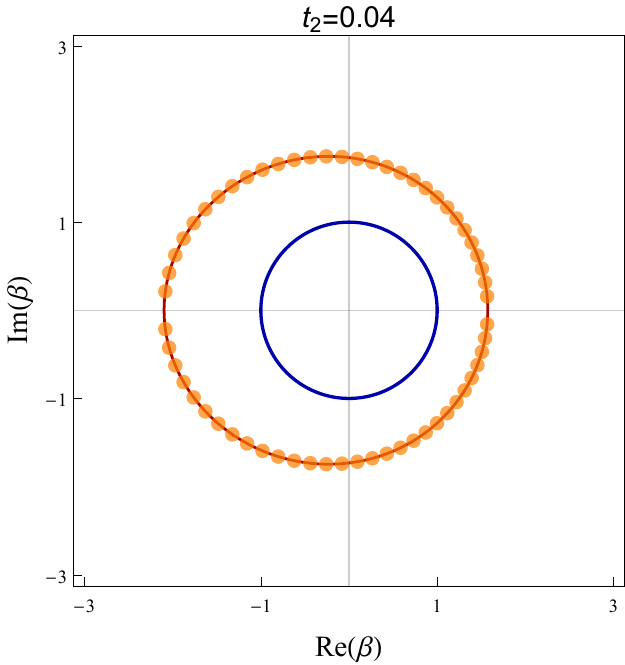}
	\hspace*{0.03\textwidth}
	\includegraphics[width=0.28\linewidth]{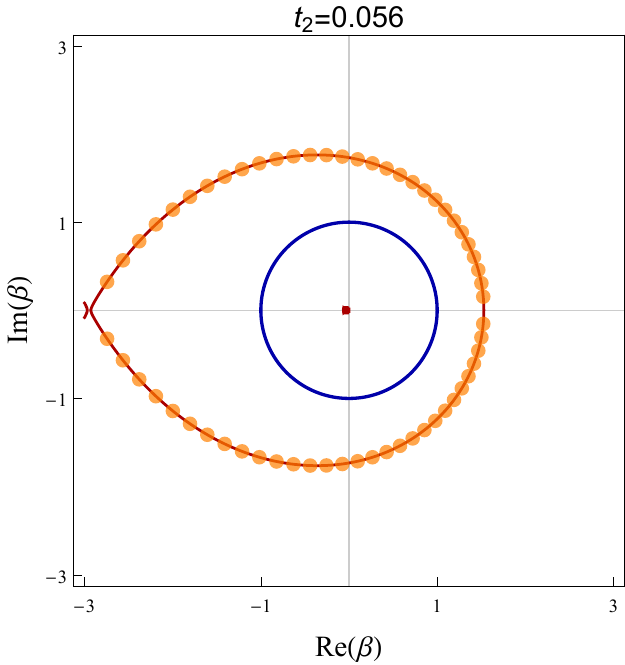}
	\includegraphics[width=0.36\linewidth]{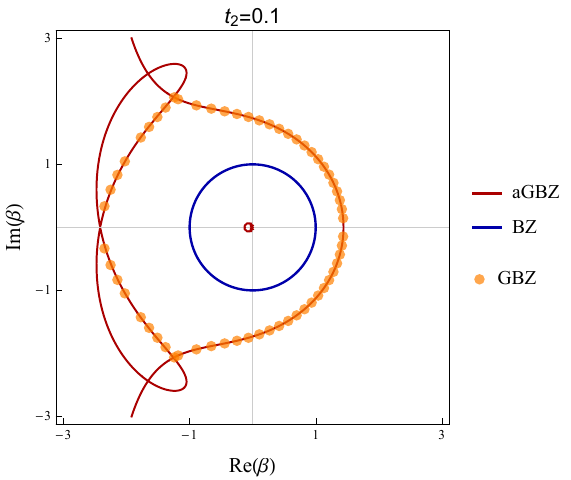}\\
	\hspace*{-0.5\textwidth}
	\includegraphics[width=0.5\linewidth]{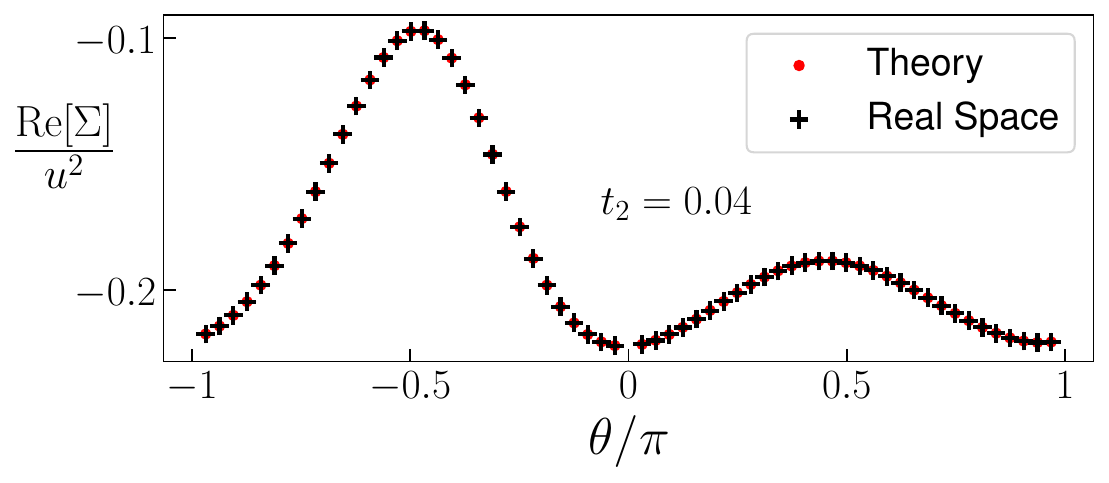}
	\includegraphics[width=0.5\linewidth]{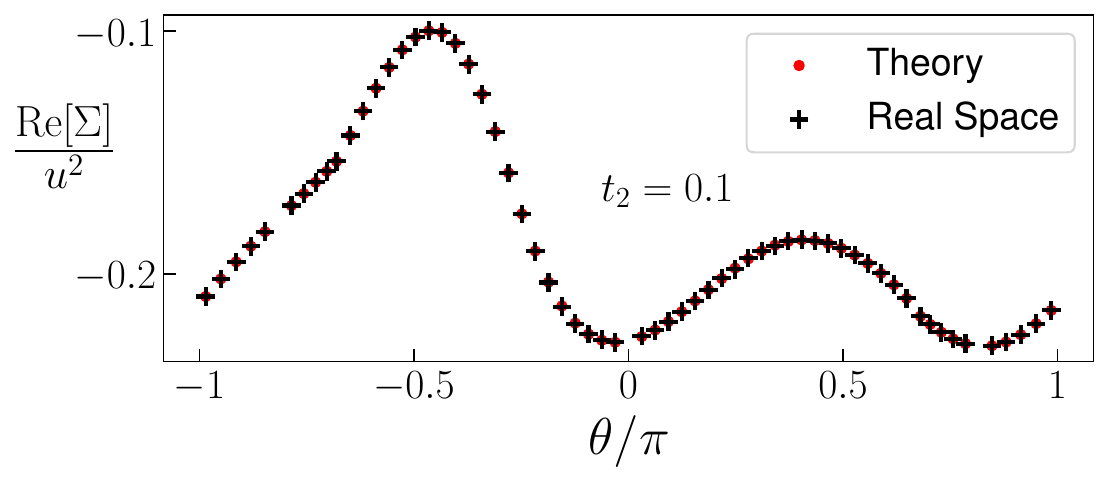}
	
	\caption{First row: generalize Brillouin zone for various values of $t_2$. Models correspond to the PT-symmetric phase, the PT-transition point, and the PT-broken phase, respectively. Red curves denote auxiliary GBZ (aGBZ, Ref. \cite{yang2020Auxiliary}); orange points (GBZ) are obtained through exact diagonalizations of finite-size systems; BZ is plotted in blue for comparison. Second row: self-energy corrections as a function of angle of $\beta$ on GBZ. $t_2=0.04$ and $0.1$.}
	\label{aGBZ}
\end{figure}

\section{Numerical results of self-energy for other models}
We apply the non-Bloch self-energy formula to the model with next-nearest-neighbor hopping of strength $t_2$, for which the analytical integral on the GBZ meets challenges. The Laurent polynomial of damping matrix is $X(\beta)=-\gamma_0-i(t-\gamma)\beta-i(t+\gamma)\beta^{-1}-i t_2(\beta^2+\beta^{-2})$. Turning on $t_2$ from $0$, when it is smaller than a critical value, the spectrum of the damping matrix remains being a straight line accompanying with the GBZ deviating from the circle on the complex plane. After $t_2$ approaches $t_{2c}$, cusps appear on GBZ, and the spectrum undergoes PT-symmetry breaking and develops complex eigenvalues. We take $t=1,\gamma=0.5, \gamma_0=1.1$, such that $t_{2c}=0.056$ which can be decided using the method introduced by Ref.~\cite{Hu2023Geometric}. The deformation of the GBZ as $t_2$ changes is shown in the first row of Fig.~\ref{aGBZ}. In the second row, we compare the self-energy obtained from the real-space effective Hamiltonian and the integral formula on BZ, which match really well.

\section{More results about GBZ deformation}

In the main text, we have presented the GBZ deformation for various values of $\gamma$ in the Liouvillian Hatano-Nelson model. In this section, we report additional parameter sweeps. We slightly extend the model by including onsite quantum jump operators: $J_j^l = \sqrt{\frac{\gamma_0}{2}}c_j$, $J_j^g = (J_j^l)^\dagger$. This modification adds a constant to the Laurent polynomial of damping matrix: $X(\beta)=-2\gamma-\gamma_0-i(t-\gamma)\beta-i(t+\gamma)\beta^{-1}$, while leaving eigenstates unchanged.

\begin{figure}[h]
	\hspace*{-0.5\textwidth}
	\includegraphics[width=0.46\linewidth]{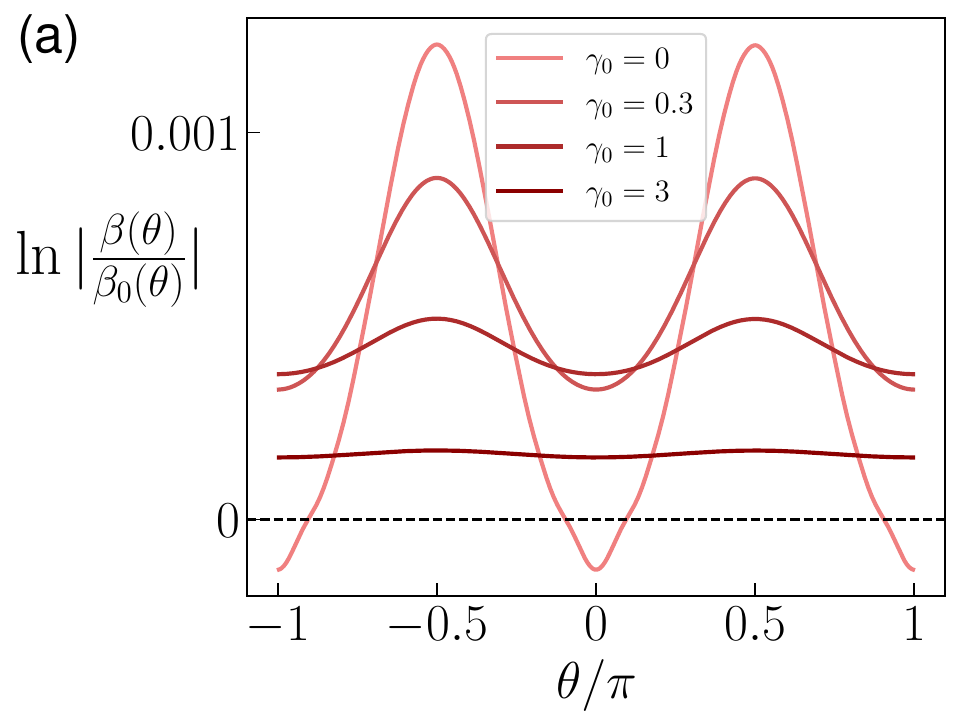}
	\includegraphics[width=0.5\linewidth]{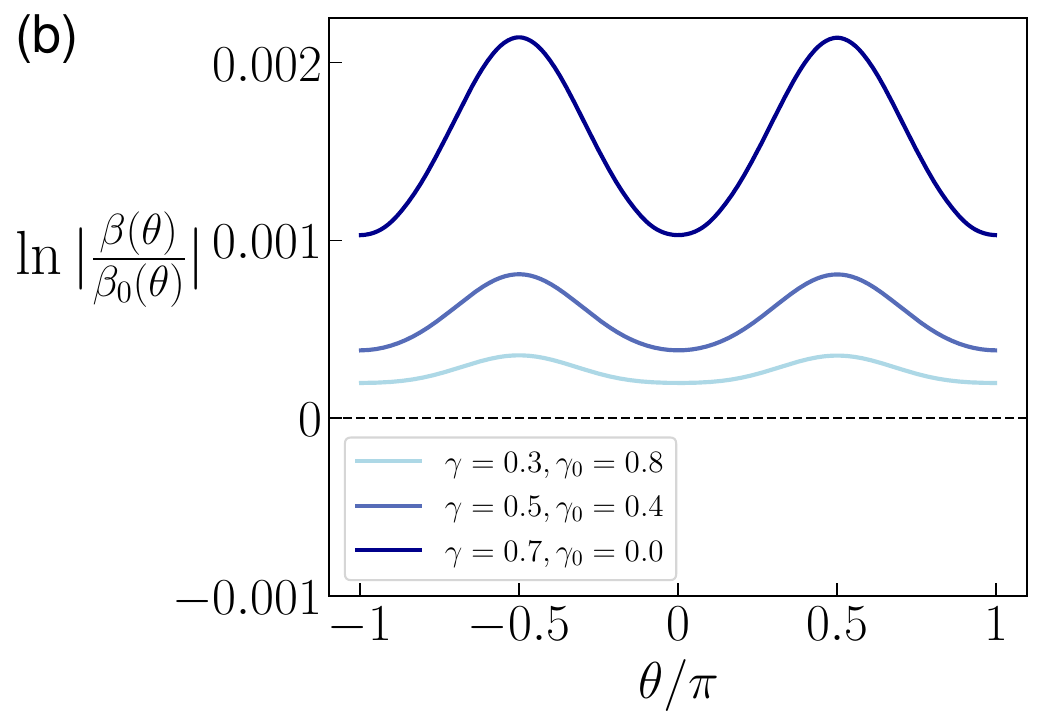}
	\caption{Angular-dependence of the relative deformation of the generalized Brillouin zone induced by
interactions. (a) $t=1, \gamma=0.5, u=0.2$, varying $\gamma_0$. (b) Fixing the Liouvillian gap $g=\gamma_0 + 2\gamma = 1.4$, $u=0.2$, varying $\gamma$ and $\gamma_0$ accordingly. }
	\label{fig:more_on_deform}
\end{figure}

Fig.~\ref{fig:more_on_deform}(a) shows the relative change of $|\beta(\theta)|$ for several values of $\gamma_0$, with $t=1$, $\gamma=0.5$, and $u=0.2$. As $\gamma_0$ increases, the angular oscillation of relative change diminishes, leading to a restoration of an isotropic GBZ.

This can be understood in the large-$\gamma_0$ limit analysis of the non-Bloch self-energy [Eq.~(7) in the main text]. Since the GBZ of this model is a circle of radius $R=\sqrt{\frac{t+\gamma}{t-\gamma}}$, we parameterize the complex momenta as follows:
\begin{equation}
    \beta = R e^{i\theta},~E = X(\beta) = -\gamma_0-2\gamma-2i\sqrt{t^2-\gamma^2}\cos(\theta),~\beta_{1,2,3} = R e^{i\theta_{1,2,3}}.
\end{equation}
We expand the self-energy as series of the inverse Liouvillian gap $1/g$ with $g = 2\gamma+\gamma_0$:
\begin{align}
    \Sigma^{(2)}(X(\beta),\beta)
=&-\frac{1}{4}\int_0^{2\pi}\frac{d\theta_{1}}{2\pi}\int_0^{2\pi}\frac{d\theta_{2}}{2\pi}\int_0^{2\pi}\frac{d\theta_{3}}{2\pi}\sum_r\frac{e^{ir(\theta_1+\theta_2+\theta_3)}}{\beta^r} 
\frac{R^3[U(R^2e^{i\theta_2+i\theta_3})^2-U(R^2e^{i\theta_2+i\theta_3})U(R^2e^{i\theta_1+i\theta_3})]}{4\gamma+2\gamma_0+2i\sqrt{t^2-\gamma^2}[\cos(\theta_1)+\cos(\theta_2)-\cos(\theta_3)-\cos(\theta)]}\nonumber\\
=&-\frac{u^2R^3}{4g}\nonumber\\
&+\frac{iR^2\sqrt{t^2-\gamma^2}}{8g^2}\int_0^{2\pi}\frac{d\theta_{1}}{2\pi}\int_0^{2\pi}\frac{d\theta_{2}}{2\pi}\int_0^{2\pi}\frac{d\theta_{3}}{2\pi}\sum_r\frac{e^{ir(\theta_1+\theta_2+\theta_3)}}{\beta^r} [U(R^2e^{i\theta_2+i\theta_3})^2-U(R^2e^{i\theta_2+i\theta_3})U(R^2e^{i\theta_1+i\theta_3})]\nonumber\\
&\times [\cos(\theta_1)+\cos(\theta_2)-\cos(\theta_3)-\cos(\theta)]\nonumber\\
&+ O(\frac{1}{g^3}).
\end{align}
The leading contribution of order $O(1/g)$ is independent of $\beta$ and therefore does not alter GBZ. The second leading term of order $O(1/g^2)$ is a Laurent polynomial in $\beta$, where the coefficient of $\beta^{-r}$ is given by the $(r,r,r)$ Fourier coefficient of a harmonic function of $\theta_{1,2,3}$. Given the form of interaction $U(\beta) = u(\beta+\beta^{-1})$, only $r=0, \pm 1$ components are nonzero, calculated respectively as follows:
\begin{equation}
    r=0: -\frac{iR^2\sqrt{t^2-\gamma^2}}{8g^2}\int_0^{2\pi}\frac{d\theta_{1}}{2\pi}\int_0^{2\pi}\frac{d\theta_{2}}{2\pi}\int_0^{2\pi}\frac{d\theta_{3}}{2\pi} U(R^2e^{i\theta_2+i\theta_3})^2\cos(\theta) = -\frac{iu^2R^2\sqrt{t^2-\gamma^2}}{8g^2}\cos(\theta),
\end{equation}
\begin{equation}
    r=1: \frac{iR^2\sqrt{t^2-\gamma^2}}{8g^2}\int_0^{2\pi}\frac{d\theta_{1}}{2\pi}\int_0^{2\pi}\frac{d\theta_{2}}{2\pi}\int_0^{2\pi}\frac{d\theta_{3}}{2\pi}\frac{e^{i(\theta_1+\theta_2+\theta_3)}}{\beta} U(R^2e^{i\theta_2+i\theta_3})U(R^2e^{i\theta_1+i\theta_3})\cos(\theta_3) = \frac{iu^2(t-\gamma)}{16g^2}\beta^{-1},
\end{equation}
and
\begin{equation}
    r=-1: \frac{iR^2\sqrt{t^2-\gamma^2}}{8g^2}\int_0^{2\pi}\frac{d\theta_{1}}{2\pi}\int_0^{2\pi}\frac{d\theta_{2}}{2\pi}\int_0^{2\pi}\frac{d\theta_{3}}{2\pi}e^{-i(\theta_1+\theta_2+\theta_3)}\beta U(R^2e^{i\theta_2+i\theta_3})U(R^2e^{i\theta_1+i\theta_3})\cos(\theta_3) = \frac{iu^2(t+\gamma)}{16g^2}\beta.
\end{equation}
Adding corrections to the non-Bloch self-energy, we obtain
\begin{equation}
    X(\beta) + \Sigma^{(2)}(X(\beta),\beta) = \text{Const.} -i (t+\gamma - \frac{u^2(t-\gamma)}{16g^2})g^{-1} - i (t-\gamma - \frac{u^2(t+\gamma)}{16g^2})\beta + O(\frac{1}{g^3}).
\end{equation}
Up to the order of $O(1/g^2)$, the deformed GBZ remains a circle, with the relative change of radius a positive value approximated as 
\begin{equation}
    \ln|\frac{\beta(\theta)}{\beta_0(\theta)}| = \ln\sqrt{\frac{(t+\gamma - \frac{u^2(t-\gamma)}{16g^2})(t-\gamma)}{(t-\gamma - \frac{u^2(t+\gamma)}{16g^2})(t+\gamma)}}\approx \frac{u^2 t\gamma}{8g^2(t^2+\gamma^2)} >0 .
\end{equation}
This shows that the angular oscillations decay polynomially in the inverse Liouvillian gap $g$ (equivalently, in $\gamma_0$), while the relative change remains positive, implying the enhancement of NHSE.

Through the above analysis, we have demonstrated the role of Liouvillian gap in the interaction-induced deformation of GBZ. In Fig.~3 of the main text, varying $\gamma$ simultaneously changes the Liouvillian gap and the non-reciprocity. To isolate the effect of the latter, we fix the gap at $g=1.4$ and adjust $\gamma$ and $\gamma_0$ accordingly, as shown in Fig.~\ref{fig:more_on_deform}(b). The angular oscillations become more pronounced as the nonreciprocity increases.

\bibliographystyle{apsrev4-1-title}
\bibliography{Heran_Feynman}